\documentclass[11pt]{article}

\usepackage{jheppub}
\usepackage{wasysym}

\usepackage{tabularx}
\usepackage{physics}
\usepackage{tikz}

\usepackage{graphicx} 
\usepackage{caption}
\usepackage{subcaption}

\usepackage{ae}

\usepackage{tabularx}

\usepackage{graphicx}

\usepackage{amsfonts,amsmath,amsthm,amssymb,amsbsy}

\usepackage{mathtools}
\usetikzlibrary{calc}

\newcommand{\dlog}{\dd \hspace{-0.07cm}\log}
\newcommand{\Dlog}{\hspace{0.02cm}\mathfrak{D}\hspace{-0.06cm} \log}

\title{Canonical Differential Equations for Cosmology from Positive Geometries}
\author[]{Mattia Capuano,}\emailAdd{m.capuano@herts.ac.uk}
\author[]{Livia Ferro,}\emailAdd{l.ferro@herts.ac.uk}
\author[]{Tomasz \L ukowski,}\emailAdd{t.lukowski@herts.ac.uk}
\author[]{and Alessandro Palazio}\emailAdd{a.palazio@herts.ac.uk}

\affiliation[]{Department of Physics, Astronomy and Mathematics, \\ University of Hertfordshire, \\  Hatfield, Hertfordshire, AL10 9AB, United Kingdom}

\abstract{
Cosmological correlation functions are central observables in modern cosmology, as they encode properties of the early universe.
In this paper, we derive novel canonical differential equations for  wavefunction coefficients in power-law FRW cosmologies by combining positive geometries and the combinatorics of tubings of Feynman graphs. First, we establish a general method to derive differential equations for any function given as a twisted integral of a logarithmic differential form. By using this method on a natural set of functions labelled by tubings of a given Feynman diagram, we derive a closed set of differential equations in the canonical form. The coefficients in these equations are related to region variables with the same notion of tubings, providing a uniform combinatorial description of the system of equations.  
We provide explicit results for specific examples and conjecture that this approach works for any graph.}

\begin{document}

\maketitle


\section{Introduction}

In the past decade, we have seen an explosion of applications of novel geometric and combinatorial methods to calculate, or bootstrap, quantities in quantum field theories, conformal field theories and cosmology. These methods rely on the notion of positivity, and they are set up directly in the kinematic space of the problem at hand, allowing for direct calculations of observables without referring to any underlying local description of the theory given by the action principle. The usual lore is to capture the singularities and factorisation properties of observables by encoding them in geometry as the locations and the factorisation properties of boundaries of a particular region in the kinematic space. These geometries are usually referred to as positive geometries \cite{Arkani-Hamed:2017tmz}, with prominent examples being the amplituhedron \cite{Arkani-Hamed:2013jha} and the momentum amplituhedron \cite{Damgaard:2019ztj,Ferro:2022abq} that encode the tree-level and integrands of the (planar) $\mathcal{N}=4$ sYM scattering amplitudes. Every positive geometry is naturally equipped with a canonical differential form that diverges logarithmically at its boundaries, and from which the physical quantities can be easily extracted. Despite being defined to all loop orders in perturbation theory, explicit calculations with amplituhedra are usually quite difficult due to the fact that they are carved out by polynomial inequalities, making the geometry very intricate, and the boundary structure complex. However, it is usually the combinatorics that comes to the rescue and allows for a classification of the boundary stratifications of these spaces, providing a natural labelling for physical quantities. For amplituhedra, the combinatorial description is closely related to the one of the positive Grassmannian spaces, whose combinatorics is well understood.

In recent years, similar methods have been applied to cosmology. The positive geometries that are relevant for cosmological observables are, however, much simpler in nature. The quantity that one calculates in cosmological models is the so-called wavefunction of the universe, and in the flat-space cosmology the relevant positive geometry is the cosmological polytope \cite{Arkani-Hamed:2017fdk,Arkani-Hamed:2018bjr,Benincasa:2024leu}. This is defined by linear inequalities, making the all-loop calculations much more feasible when compared to the ones for the $\mathcal{N}=4$ sYM scattering amplitudes. Also in this case a combinatorial description of these positive geometries exists, and it is given by tubes and tubings associated with Feynman graphs contributing to the perturbative expansion of cosmological wavefunctions. For a given Feynman graph, finding all its tubings is an easy task, which allows one to write an explicit form of the flat-space wavefunctions to a very high order in perturbation theory.

These methods also allow for finding quantities in cosmological theories beyond flat space. Most work has been done in certain toy models, with the most prominent role played by the theory of conformally-coupled scalar fields in power-law Friedmann-Robertson-Walker (FRW) cosmologies with polynomial interactions. Importantly, the wavefunctions in the FRW cosmologies can be found by performing particular (twisted) integrals of the flat-space results. Motivated by similar problems in the Feynman integrals community, various sets of differential equations satisfied by the FRW wavefunctions have been suggested in recent years, with the twisted cohomology \cite{De:2023xue,De:2024zic,Gasparotto:2024bku}, integration-by-parts \cite{Chen:2023iix,Chen:2024glu} and the kinematic flow approaches \cite{Arkani-Hamed:2023kig,Arkani-Hamed:2023bsv,Baumann:2024mvm,Baumann:2025qjx} providing the most comprehensive answers (see also \cite{Fan:2024iek,Grimm:2024mbw,Grimm:2024tbg,Grimm:2025zhv,He:2024olr,Benincasa:2024ptf,Hang:2024xas,Fevola:2024nzj} and references therein).  
In particular, the recent results in \cite{Baumann:2025qjx} provide a physics-motivated basis of integrals for the differential equations, originating from a natural decomposition of the bulk-to-bulk propagator in FRW cosmologies into a time-ordered piece, an anti-time-ordered piece, and a non-time-ordered piece. The elements of this basis have a combinatorial description in terms of graph tubings for particular dressed graphs. While these dressed graphs properly capture the combinatorics of the kinematic flow, they significantly differ from the natural tubings for graphs used in the definition of the cosmological polytopes. 

Our goal in this paper is to significantly extend these recent developments. We will provide a different set of differential equations for the FRW wavefunctions, which is directly rooted in the combinatorics of the graph tubings and is naturally motivated by positive geometries. Our choice of functions will rely on the relation recently discussed in \cite{Glew:2025ugf} between the wavefunction coefficients with the amplitubes introduced in \cite{Glew:2025otn}.  In our approach we insist on staying inside the space of logarithmic differential forms, and provide a simple algorithm to derive the differential equations, utilising simple properties of such differential forms. Importantly, the resulting differential equations are automatically in the canonical form. Our set of functions is in one-to-one correspondence with particular tubings on a given Feynman graph, and the differential equations coefficients are given by region variables that are naturally defined in terms of these tubings. This allows for a more uniform combinatorial description of the result, and avoids introducing any dressing of graphs. Importantly, since we insist on staying in the realm of logarithmic forms, the number of functions that we need to include in our construction is larger than the one in the kinematic flow approach, and is counted by the number of graph tubings. Finally, the matrix of coefficients in the equations that we derive in this paper can be diagonalised  very easily, and the resulting set of eigenvectors and eigenvalues can be again interpreted using the same notion of graph tubings.

This paper is organised as follows: in Section \ref{sec:wavefunctionsandtubings} we set up the stage by reviewing the cosmological quantities that we consider in this paper, and by providing the ways to calculate them using geometry and combinatorics. In particular, we introduce the definition of graph tubings for connected graphs. In Section \ref{sec:diffeqs} we derive the differential equations satisfied by the cosmological wavefunction coefficients, utilising the properties of logarithmic differential forms and of graph tubings. Afterwards, in Section \ref{sec:examples} we give a very detailed explanation on how our method works in a few basic examples.  We end the paper with conclusions and an outlook on interesting directions opened by this work.


\section{Cosmological Wavefunctions and Tubings}\label{sec:wavefunctionsandtubings}
\subsection{Wavefunctions in FRW Cosmologies}

In this paper we will focus on an important class of cosmological toy models, which  provide a useful playground to understand properties of more realistic theories. This setup follows closely the one explained in \cite{Baumann:2024mvm}.
In particular, we consider a conformally-coupled scalar field $\phi$, with non-conformal polynomial self-interactions, in  a FRW cosmology in four spacetime dimensions:
\begin{equation}
\label{eq:action}
S = - \int d^4x \sqrt{-g} \left(\frac{1}{2} (\partial_\mu \phi)^2+\frac{1}{12}R \phi^2+\sum_{k>2}\frac{\lambda_k}{k!}\phi^k \right),
\end{equation}
where $R$ is the Ricci scalar.
We compute correlation functions in a universe evolving as a power law in conformal time $\eta$, with scale factor $a(\eta)=\left(\frac{\eta}{\eta_0}\right)^{-(1+\epsilon)}$, where $\epsilon$ and $\eta_0$ are constant parameters. Depending on the value of the cosmological parameter $\epsilon$ we recover different cosmologies, such as de Sitter $(\epsilon=0)$, flat space $(\epsilon=-1)$, or related to inflation $(\epsilon \sim 0)$.
These power-law cosmologies are particularly interesting because the action \eqref{eq:action} can be related to the one of a massless field theory in  flat space with time-dependent couplings, which in turn allows to simplify the computation of correlation functions.

\begin{figure}
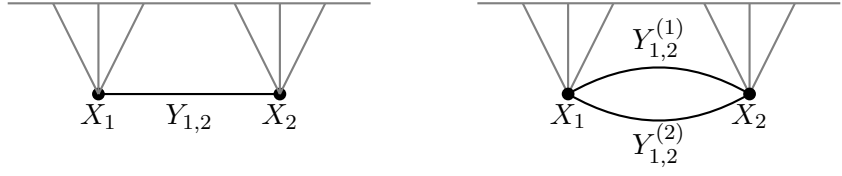

\begin{center}
\raisebox{0.45cm}{\input{Figures/Feynman1.tex}}
\hspace{1cm}
\input{Figures/Feynman2.tex}
\end{center}
\caption{Examples of Feynman graphs for wavefunction coefficients at tree level (left) and one loop (right).}\label{fig:feynman}
\end{figure}

We study correlation functions at a fixed conformal time $\eta=\eta_*=0$ which are calculated as functional integrals
\begin{equation}
\langle \varphi(\mathbf{x}_1)\ldots  \varphi(\mathbf{x}_N)\rangle=\int \mathcal{D}\varphi\,\varphi(\mathbf{x}_1)\ldots  \varphi(\mathbf{x}_N) |\Psi[\varphi]|^2\,,
\end{equation}
where $\varphi(\mathbf{x})$ is the boundary value of the field $\phi(\eta,\mathbf{x})$, $\varphi(\mathbf{x}) \equiv \phi(0,\mathbf{x})$. Here, $\Psi[\varphi]$ is the wavefunction and it admits an expansion in the Fourier space of the form
\begin{equation}
\Psi[\varphi]=\exp\left[-\sum_{n=2}^\infty \int \left(\prod_{a=1}^n\frac{\dd^3 k_a}{(2\pi)^3}\varphi_{\mathbf{k}_a}\right)\psi_n(\mathbf{k}_1,\ldots,\mathbf{k}_n) (2\pi)^3 \delta^3(\mathbf{k}_1+\ldots \mathbf{k}_n)\right] .
\end{equation}
 The functions $\psi_n(\mathbf{k}_1,\ldots,\mathbf{k}_n)$ are referred to as the wavefunction coefficients and they depend on $n$ three-vectors ${\bf{k}}_a$, the spatial momenta, whose magnitudes $|{\bf{k}}_a|$ we refer to as ``energies". The wavefunction coefficients  can be represented as sums of Feynman diagrams and therefore computed using Feynman rules, similar to scattering amplitudes in quantum field theories; we illustrate particular contributions at tree- and loop-level in Fig. \ref{fig:feynman}.
To find Feynman graphs for a given $n$, one draws all graphs with $n$ lines ending on the spatial surface at $\eta=\eta_*= 0$. To each external line one associates a bulk-to-boundary propagator and for each internal line one assigns a bulk-to-bulk propagator. 
Finally, one integrates over all vertex times with measure $d\eta \,a^{4}(\eta)$, and over all undetermined loop momenta. 
For conformally-coupled fields, the function associated to a Feynman graph $G$  depends only on the variables $X_v$ that are the sums of the external energies entering each vertex $v$, and on $Y_e$ that are the energies associated to each internal edge $e$ of the graph. Therefore the number of external lines is irrelevant and they can be amputated, leaving one with the so-called reduced graph.

Importantly, any FRW wavefunction coefficient for generic $\epsilon$ can be constructed from their flat-space counterparts ($\epsilon=-1$), which in turn have a geometric description in terms of the cosmological polytope \cite{Arkani-Hamed:2017fdk}. More precisely, the cosmological wavefunction in a power-law FRW spacetime is obtained from the flat-space wavefunction by shifting the external energies $X_v\rightarrow X_v+x_v$ and performing the integral over $x_v$ with the so-called twist function $u=\prod_{v} x_v^{\alpha_v}$:
\begin{equation}\label{eq:FRWintro}
\psi_G^{\text{FRW}}(X_v,Y_e)=\int_0^\infty \left(\prod_v \dd x_v \,x_v^{\alpha_v}\right) \psi_G^{\text{flat}} (X_v+x_v,Y_e) \,,
\end{equation}
where we introduced generic parameters ${\alpha_v}$, which are related to the underlying cosmology through the cosmological parameter $\epsilon$. The twisted integral \eqref{eq:FRWintro} will be the main quantity we study in this paper.

As an example, for the tree-level process in Fig. \ref{fig:feynman}, the wavefunction coefficient in flat space associated to this graph is
\begin{equation}
\label{eq:exflat}
\psi_G^{\text{flat}}=\frac{2Y_{1,2}}{(X_1+Y_{1,2})(X_2+Y_{1,2})(X_1+X_2)} \,.
\end{equation}
Then, the FRW wavefunction coefficient for the same tree-level process is simply constructed using the flat-space counterpart \eqref{eq:exflat} as:
\begin{equation}
\psi_G^{\text{FRW}}=\int_0^{\infty} \dd x_1 \dd x_2 \,x_1^{\alpha_{1}} x_2 ^{\alpha_{2}}\frac{2Y_{1,2}}{(X_1+x_1+Y_{1,2})(X_2+x_2+Y_{1,2})(X_1+x_1+X_2+x_2)} \,.
\end{equation}
In the following section, we explain in detail how to compute the flat-space wavefunction using the combinatorics of tubings on graph $G$.

\subsection{Tubings}\label{sec:tubings}

As shown in \cite{Arkani-Hamed:2017fdk}, the flat-space wavefunction $\psi^{\text{flat}}_G (X_v,Y_e)$ can be calculated combinatorially using tubes and tubings associated to the underlying graph $G$. Importantly, there are three types of graph tubings that will be relevant to this paper: binary tubings ($b$-tubings), unary tubings ($u$-tubings), and cut tubings ($c$-tubings). We revisit their definition here, closely following the exposition in \cite{Glew:2025ugf}, and provide some additional properties that will be relevant to the latter parts of this paper. 

For a given graph $G$, we will denote the set of its vertices by $V_G=\{v_1,\ldots,v_{|V_G|}\}$, and the set of its edges by $E_G=\{e_{i,j}\}$, where the edge $e_{i,j}$ is incident to vertices $v_i$ and $v_j$. We allow graphs with multiple edges between vertices, and if more than one edge connects a pair of vertices, then we will add an additional label $e_{i,j}^{(k)}$ to distinguish them. In the following we will focus mostly on connected graphs; however, many definitions that we present here can be easily extended to a generic graph.

\begin{figure}[t]
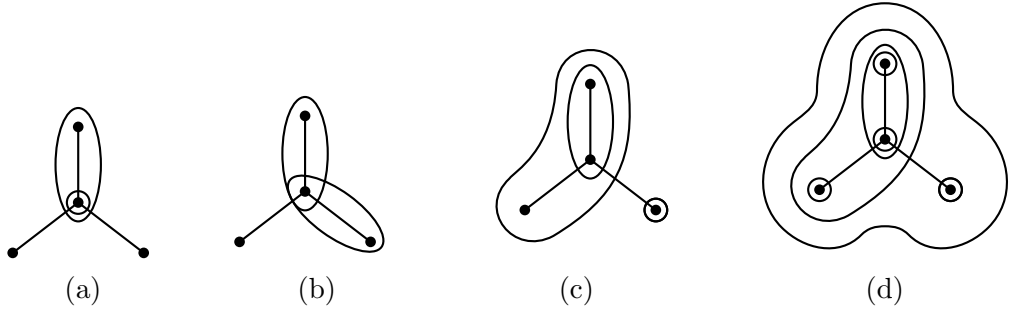

 \begin{center}
 \begin{tabular}{cccc}
 \input{Figures/btubings1.tex}\quad\phantom{a}& \input{Figures/btubings2.tex}\quad\phantom{a}&\input{Figures/btubings3.tex}\quad\phantom{a}&\input{Figures/btubings4.tex}\\
 (a)\quad\phantom{a}&(b)\quad\phantom{a}&(c)\quad\phantom{a}&(d)
 \end{tabular}
 \end{center}
 \caption{An example of $b$-tubes that are (a) compatible and (b) not compatible on the star graph with four vertices. An example of (c) a $b$-tubing and (d) a maximal $b$-tubing.}
 \label{fig:btubings}
 \end{figure}

Given a connected graph $G$ we define a {\it $b$-tube} $T$ as a connected subgraph of $G$ and we denote the set of all $b$-tubes on $G$ as $B_G$. To each $b$-tube $T$ we can associate a linear function defined as\footnote{The definition provided in this paper is phrased differently compared to the previous literature, and does not use the notion of ``an edge cuts a tube'' or ``an edge enters a tube''. We believe that our definition of $H_T$ is more precise.}
\begin{equation}\label{eq:Hb}
H_T=\sum_{v\in V_T}\left(X_v+\sum_{e=[v,v'] \in G\setminus T} Y_e\right),
\end{equation}
where the second sum runs over all edges incident to $v$ in $G$ that are in the complement of $T$, and self-loops $e=(v,v)$ are counted twice. In this formula we introduced the variables $X_v$ for each vertex $v\in V_G$ and the variables $Y_e$ for each edge  $e\in E_G$. 
 We say that two $b$-tubes $T_1$ and $T_2$ are {\it compatible} if either one of them is a subset of the other ($T_1\subset T_2$ or $T_2\subset T_1$), or if their vertex sets are disjoint ($V_{T_1}\cap V_{T_2}=\emptyset$). A collection of $b$-tubes $\mathbf{T}\subset B_G$ forms a {\it $b$-tubing} if all the tubes in $\mathbf{T}$ are mutually compatible. A $b$-tubing is called maximal if no more compatible $b$-tubes can be added to it. We denote the set of all $b$-tubings by $\mathcal{B}_G$, and the set of all maximal $b$-tubings as $\mathcal{B}_G^{\text{max}}$. For a given tubing $\mathbf{T}\in \mathcal{B}_G$ we define $n_\mathbf{T}=|\mathbf{T}|$ to be the number of tubes in $\mathbf{T}$. Importantly, for each  maximal $b$-tubing $\mathbf{T}\in\mathcal{B}^{\text{max}}_G$ we have $n_\mathbf{T}=|V_G|+|E_G|$. We illustrate the notions of $b$-tubes and $b$-tubings in Fig.~\ref{fig:btubings}, where we indicated a $b$-tube $T$ in the graph drawing by encircling the vertices and edges of the graph. 
 
\begin{figure}[t]
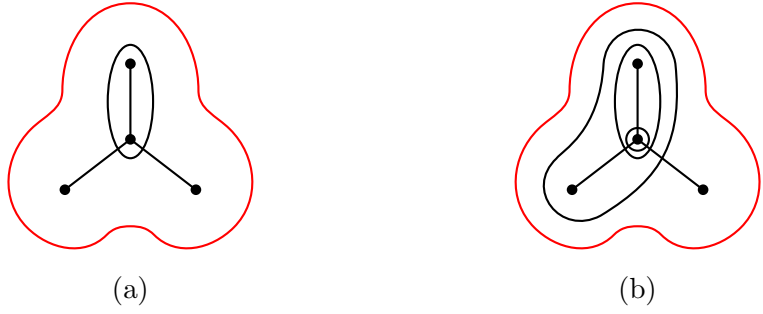

 \begin{center}
 \begin{tabular}{ccc}
 \input{Figures/utubings1.tex}&\phantom{a}\hspace{2cm}\phantom{a}& \input{Figures/utubings2.tex}\\
 (a)&&(b)
 \end{tabular}
 \end{center}
 \caption{An example of (a) a $u$-tubing and (b) a maximal $u$-tubing on the star graph with four vertices.}
 \label{fig:utubings}
 \end{figure}

A different notion of tubes and tubings relevant to this paper was introduced in \cite{CARR20062155}. For a connected graph $G$, we define a {\it $u$-tube} $T$ on $G$ as the graph induced by a non-empty subset of vertices $V\subset V_G$, such that $T$ is a connected graph. When the graph $G$ is fixed, specifying a $u$-tube is equivalent to specifying the vertex subset $V$, and therefore we can label any $u$-tube as $T=\{v_{i_1},\dots,v_{i_{|V_T|}}\}\subset V_G$. We give special importance to the $u$-tube $T=V_G$ that contains all vertices of graph $G$ and call it the {\it root}. We denote the set of all $u$-tubes on $G$ as $U_G$. We notice that every $u$-tube is trivially a $b$-tube and therefore $U_G\subset B_G$. We say that two $u$-tubes $T_1$ and $T_2$ are compatible if they do not intersect (namely $T_1\cap T_2=\emptyset$ or $T_1\subset T_2$ or $T_2\subset T_1$) and they are not adjacent (namely if $T_1\cap T_2=\emptyset$ then $T_1 \cup T_2\notin U_G$). We define a {\it $u$-tubing} $\mathbf{T}$ as a collection of  $u$-tubes that are mutually compatible and that contain the root $V_G\in\mathbf{T}$, and a maximal $u$-tubing is a maximal set of mutually compatible $u$-tubes. In the following, when depicting the $u$-tubings, we will indicate the root in red, while $u$-tubes other than the root will be indicated in black. The set of $u$-tubings will be denoted by $\mathcal{U}_G$ and the set of maximal $u$-tubings will be denoted as $\mathcal{U}_G^{\text{max}}$. The number of $u$-tubes in any maximal $u$-tubing is $|V_G|$. An example of $u$-tubing and a maximal $u$-tubing can be found in Fig.~\ref{fig:utubings}.

\begin{figure}[t]
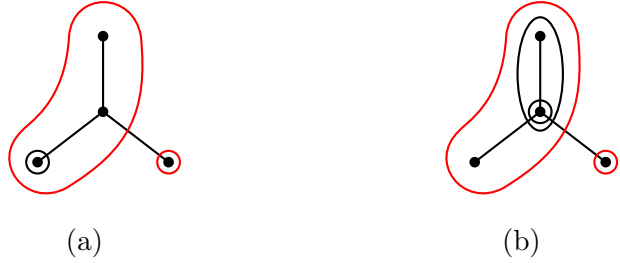

 \begin{center}
 \begin{tabular}{ccc}
 \input{Figures/ctubings1.tex}&\phantom{a}\hspace{2cm}\phantom{a}& \input{Figures/ctubings2.tex}\\
 (a)&&(b)
 \end{tabular}
 \end{center}
 \caption{Two examples of $c$-tubings on the star graph with four vertices. The tubing in figure (b) is maximal, while the one in figure (a) is not.}
 \label{fig:ctubings}
 \end{figure}

Finally, we define {\it $c$-tubings} on $G$ as a particular mixture of $b$-tubes and $u$-tubes. To define a $c$-tubing $\tau$, we select a subset of edges of $G$, $\mathbf{e}\subseteq E_G$. The graph $G_\mathbf{e}=G[E_G\setminus \mathbf{e}]$ that is obtained from $G$ by removing the edges in $\mathbf{e}$ is not necessarily connected, and we denote the set of its connected components by $\mathcal{CC}(G_{\mathbf{e}})=\{T_1,\ldots,T_p\}$. Importantly, each connected component $T_i\in\mathcal{CC}(G_{\mathbf{e}})$ is a $b$-tube on $G$, and $T_i$ and  $T_j$ are compatible $b$-tubes for $i\neq j$, which means that the set $\mathbf{e}$ naturally defines a $b$-tubing of $G$. The $b$-tubes $T_i$ in this $b$-tubing will play the role of the root $u$-tubes for the corresponding connected component. Then for each graph $T_i\in\mathcal{CC}(G_{\mathbf{e}})$ we take any $u$-tubing $\tau_i$ of $T_i$. A $c$-tubing $\tau$ is then defined as the union
\begin{equation}
\tau=\tau_1\cup\ldots\cup\tau_k\,.
\end{equation}
A $c$-tubing is called maximal if all $u$-tubings $\tau_i$ are maximal. Again, the number of tubes in a maximal $c$-tubing is $|V_G|$. We will denote the set of all $c$-tubings on $G$ as $\mathcal{C}_G$, and the set of all maximal $c$-tubings as $\mathcal{C}_G^{\text{max}}$. We note that the set of $u$-tubings is a subset of $c$-tubings, namely every $u$-tubing of $G$ is a $c$-tubing of $G$ for $\mathbf{e}=\emptyset$, and therefore $\mathcal{U}_G\subset\mathcal{C}_G$. Two examples of $c$-tubings on the star graph with four vertices are depicted in Fig.~\ref{fig:ctubings}. 

\begin{figure}[t]
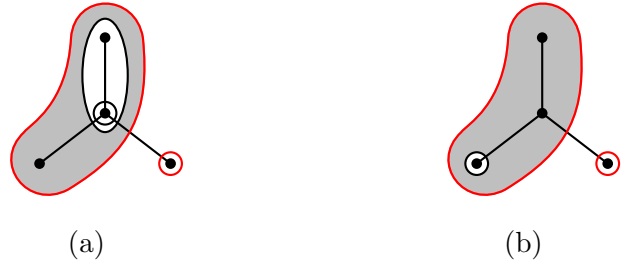

 \begin{center}
 \begin{tabular}{ccc}
 \input{Figures/region1.tex}&\phantom{a}\hspace{2cm}\phantom{a}& \input{Figures/region2.tex}\\
 (a)&&(b)
 \end{tabular}
 \end{center}
 \caption{Two regions on the star graph with four vertices. The shaded areas are bounded by the parent tube of each region, and they do not include areas bounded by the children tubes.}
 \label{fig:regions}
 \end{figure}

For a given $c$-tubing on graph $G$, we define {\it regions} as follows: for a $c$-tubing $\tau=\{T_1,\ldots,T_p\}$, we define a partial ordering on the set of its tubes by inclusion. Then the region $r=\{T_{i_1};T_{i_2},\ldots , T_{i_{|r|}}\}\subset\tau$ is defined as a collection of tubes such that all tubes $T_{i_j}$ are subsets of $T_{i_1}$, called the {\it parent tube}, and all other tubes, called the {\it children tubes}, are disjoint. For basic graphs, we will indicate a graph region by coloring the space between the parent tube and the children tubes, as depicted in Fig.~\ref{fig:regions}.
We will denote by $\mathcal{R}_G$ the set of all possible regions for a given graph $G$. For a given graph $G$ and its $c$-tubing $\tau$, we define $r_v$ for each $v\in V_G$ as the region containing vertex $v$, i.e.~$v$ is a vertex of the parent tube and is not a vertex of any of the children tubes in $r$. Also, for a given tube $T\in\tau$, we define two regions: $r_{\tau,T}^{\text{out}}$ as the region for which $T$ is one of the children tubes, and  $r_{\tau,T}^{\text{in}}$ as the region for which $T$ is the parent tube. Graphically, this corresponds to the regions that are adjacent to the tube $T$, with $r_{\tau,T}^{\text{out}}$ outside and $r_{\tau,T}^{\text{in}}$  inside the tube $T$ in the tubing $\tau$.
To a given region $r=\{T_1;T_2,\ldots , T_{|r|}\}$, we can associate a natural linear function of $X_v$ and $Y_e$ using the $H_T$ functions defined in \eqref{eq:Hb}
\begin{equation}
R_r=H_{T_1}-\sum_{j=2}^{|r|}H_{T_{j}}\,,
\end{equation}
where we subtracted the contributions of all children tubes from the contribution of the parent tube. The linear functions $R_r$ will play a crucial role in our derivation of differential equations later in this paper. In particular, the set of logarithmic differential forms $\dlog R_r$ for $r\in \mathcal{R}_{G}$ provides the complete set of coefficients of the differential equations for the graph $G$.

Finally, we note that for any $c$-tubing $\tau$ there is a natural map
\begin{equation}\label{eq:vertexbijection}
q_{\tau}: V_G \to \tau\,,
\end{equation}
from the set of vertices of the graph $G$ to the tubing $\tau$. To define this map we associate to a given vertex $v\in V_G$ the smallest (by inclusion) tube in $\tau$ containing $v$. This may be a $u$-tube or one of the root $b$-tubes. Therefore, to every $c$-tubing $\tau$ we can associate an ordered set of size $|V_G|$ as 
\begin{equation}
q_{\tau}(V_G)=(T_{i_1},T_{i_2},\ldots,T_{i_{|V_G|}})\,,
\end{equation}
where the tubes in this ordered set might be repeated. 
 \begin{figure}[t]
 \begin{center}

 \input{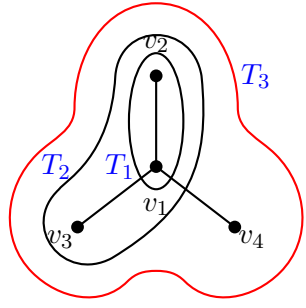}

 \end{center}
 \caption{An example illustrating the ordering of tubes obtained from the map $q_\tau$ in \eqref{eq:vertexbijection}. For the $c$-tubing $\tau=\{T_1,T_2,T_3\}$ with the tubes $T_1=\{v_1,v_2\}$, $T_2=\{v_1,v_2,v_3\}$, $T_3=\{v_1,v_2,v_3,v_4\}$ the ordering is given by $q_\tau(\{v_1,v_2,v_3,v_4\})=(T_1,T_1,T_2,T_3)$.}
 \label{fig:tubingordering}
 \end{figure}
An example of this ordering is given in Fig.~\ref{fig:tubingordering}. We will use this observation later on to define our canonical forms associated to a graph $G$, which will be in one-to-one correspondence with $c$-tubings on $G$.

\subsection{Wavefunctions and Amplitubes}

Using the notion of $b$-tubes and $b$-tubings introduced in the previous section, one can find the flat-space wavefunction coefficient $\psi_G^{\text{flat}}$ associated to a given connected graph $G$ as a sum over the maximal $b$-tubings of $G$ as \cite{Arkani-Hamed:2017fdk}
\begin{equation}\label{eq:wavefunction}
\psi_G^{\text{flat}}=\left(\prod_{e\in E_G}2Y_e\right)\sum_{\mathbf{T}\in \mathcal{B}_G^{\text{max}}}\prod_{T\in \mathbf{T}}\frac{1}{H_T} \,,
\end{equation}
where the functions $H_T$ are defined in \eqref{eq:Hb}.

Similarly, using the $u$-tubes and $u$-tubings, one defines the {\it amplitube} $A_G$ associated to a given connected graph $G$ as the sum over all maximal $u$-tubings \cite{Glew:2025otn}
\begin{equation}\label{eq:amplitube}
A_G=\sum_{\mathbf{T}\in \mathcal{U}_G^{\text{max}}}\prod_{T\in \mathbf{T}}\frac{1}{H_T}\,,
\end{equation}
where the expressions for $H_T$ are again calculated as in \eqref{eq:Hb}.

The wavefunction coefficients and amplitubes satisfy an important relation that was extensively studied in \cite{Glew:2025ugf} (see also \cite{Fevola:2024nzj})
\begin{equation}\label{eq:wavefunctionamplitube}
\psi_G^{\text{flat}}=\sum_{\mathbf{e}\subseteq E_G}(-1)^{|\mathbf{e}|}\prod_{T\in \mathcal{CC}(G_{\mathbf{e}})}A_{T}\,,
\end{equation}
where the sum runs over all subsets of the edge set $E_G$ of $G$, and the product is over amplitubes for all connected components of the graph $G_\mathbf{e}=G[E_G\setminus \mathbf{e}]$.

Importantly, both the wavefunction coefficients $\psi_G^{\text{flat}}$ and the amplitubes $A_T$ can be uplifted to canonical forms of positive geometries. In this context, the (rescaled) wavefunction coefficients are related to the canonical forms of the cosmological polytopes \cite{Arkani-Hamed:2017fdk}, while the amplitubes are related to the canonical forms of the graph associahedra in the tube embedding defined in \cite{Glew:2025otn}.
A cosmological polytope associated to a graph $G$ is a projective polytope in the space of $(X_v,Y_e)$, cut out by the inequalities 
\begin{equation}
H_T>0,\text{ for all $b$-tubes } T\in B_G\,.
\end{equation}
The canonical form $\Omega^{\text{flat}}_G$ of the resulting polytope is given by the sum over all its vertices, which in turn correspond to maximal $b$-tubings on $G$. Then
\begin{equation}\label{eq:omegacosmopoly}
\Omega^{\text{flat}}_G=\sum_{\mathbf{T}\in \mathcal{B}_G^{\text{max}}}\sigma_{\mathbf{T}}\, \dlog H_{T_1} \wedge \ldots \wedge \dlog H_{T_{|E_G|+|V_G|}}\,,
\end{equation}
where the signs $\sigma_{\mathbf{T}}$ can be fixed, up to an overall sign, by requiring that the form $\Omega^{\text{flat}}_G$ is projectively invariant, namely it is invariant under the rescaling $H_T \to \lambda H_T$ for all $T\in B_G$.
Then the relation between the wavefunction $\psi^{\text{flat}}_G$ and the canonical form of the cosmological polytope $\Omega^{\text{flat}}_G$ is given by
\begin{equation}\label{eq:omegatopsi}
\left(\prod_{e\in E_G}2Y_e\right) \Omega^{\text{flat}}_G=\psi^{\text{flat}}_G \,\dd X_1\wedge\ldots\wedge\dd X_{|V_G|}\wedge \dd Y_1\wedge\ldots\wedge\dd Y_{|E_G|}\,.
\end{equation}
Since there is a non-trivial $Y$-dependent factor in front of $\Omega^{\text{flat}}_G$ in \eqref{eq:omegatopsi}, the wavefunction does not directly descends from a logarithmic form on this space. In the following we will see that it can be rectified by considering canonical forms on a reduced space of vertex variables $X_v$, while keeping the variables $Y_e$ constant.

On the other hand, the amplitube $A_G$ \eqref{eq:amplitube} can be directly uplifted to a canonical form of a polytope, namely, it is a canonical form of the graph associahedron in the tube embedding \cite{Glew:2025otn}. We define a logarithmic differential form $\omega_G$ associated to the amplitube $A_G$ as
\begin{equation}\label{eq:upliftamplitubes}
\omega_G=\sum_{\tau\in \mathcal{U}_G^{\text{max}}}\bigwedge_{T\in \tau}\dlog H_T\,,
\end{equation}
where we keep the variables $Y_e$ constant (and therefore the differential $\dd$ only acts on the vertex variables $X_v$). We use the ordering of tubes in the $u$-tubing $\tau$ defined by the function $q_\tau$ in \eqref{eq:vertexbijection}. Then there are no additional signs required in formula \eqref{eq:upliftamplitubes}. The amplitube $A_G$ can be recovered from $\omega_G$ as the coefficient in front of the differential top form
\begin{equation}
\omega_G=A_G \,\dd X_1\wedge\ldots\wedge\dd X_{|V_G|}\,.
\end{equation}

Using the uplift of the amplitubes to differential forms \eqref{eq:upliftamplitubes}, we can now use the relation \eqref{eq:wavefunctionamplitube} to define a logarithmic form corresponding directly to the wavefunction $\psi^{\text{flat}}_G$:
\begin{equation}\label{eq:diffformsexpansion}
\widetilde{\Omega}^{\text{flat}}_G=\sum_{\mathbf{e}\subseteq E_G}(-1)^{|\mathbf{e}|}\bigwedge_{T\in \mathcal{CC}(G_{\mathbf{e}})}\omega_T=\sum_{\tau\in\mathcal{C}^{\text{max}}_G}(-1)^{|\mathbf{e}|}\mu_\tau\,,
\end{equation} 
where the former expression is a direct uplift of formula \eqref{eq:wavefunctionamplitube} to differential forms, while in the latter expression we rewrote the $\widetilde{\Omega}^{\text{flat}}_G$ as a sum over all maximal $c$-tubings on $G$, with
\begin{equation}
\mu_\tau=\dlog {H}_{T_1}\wedge\ldots\wedge \dlog {H}_{T_{|V_G|}}\,.
\end{equation}
Importantly, the differential form $\widetilde{\Omega}^{\text{flat}}_G$ differs from $\Omega^{\text{flat}}_G$ in \eqref{eq:omegacosmopoly} coming from the definition of cosmological polytope. In particular, these forms have different degrees: while $\Omega^{\text{flat}}_G$ is a $(|E_G|+|V_G|)$-form on the $(X_v,Y_e)$ space, the $\widetilde{\Omega}^{\text{flat}}_G$ form is a $|V_G|$-form on the $X_v$ space. The latter is however more directly related to the wavefunction coefficients since
\begin{equation}\label{eq:omegatildedecomp}
\widetilde{\Omega}^{\text{flat}}_G=\psi^{\text{flat}}_G \,\dd X_1\wedge\ldots\wedge\dd X_{|V_G|}\,.
\end{equation}  
In the following, we will use the differential form $\widetilde{\Omega}^{\text{flat}}_G$ as defined in \eqref{eq:diffformsexpansion} in all our calculations. Due to \eqref{eq:omegatildedecomp}, $\widetilde{\Omega}^{\text{flat}}_G$ contains all the information about the flat space wavefunction coefficients $\psi^{\text{flat}}_G$.

\subsection{From Flat-Space to Cosmological Wavefunctions}
So far we have discussed the flat-space wavefunction coefficients $\psi^{\text{flat}}_G$ associated to a given Feynman graph $G$, and their uplift to the differential form $\widetilde{\Omega}^{\text{flat}}_{G}$. As we pointed out before, the main focus of this paper is on the derivation of differential equations for cosmological wavefunction coefficients that are related to the flat-space ones by
\begin{equation}\label{eq:flattoFRW}
\psi^{\text{FRW}}_G(X_v,Y_e)=\int \left(\prod_v \dd x_v \,x_v^{\alpha_v}\right) \psi^{\text{flat}}_G (X_v+x_v,Y_e)\,.
\end{equation}
Since the arguments of the flat-space wavefunction coefficients in \eqref{eq:flattoFRW} are the linear combinations $x_v+X_v$, then it is straightforward to turn the integrand of \eqref{eq:flattoFRW} into a differential form, up to the twist function $x_v^{\alpha_v}$, as:
\begin{align}
\psi^{\text{FRW}}_G(X_v,Y_e)&=\int \left(\prod_v x_v^{\alpha_v}\right)\,\widetilde\Omega^{\text{flat}}_G (X_v+x_v,Y_e)\\\label{eq:FRWasform}
&=\int \left(\prod_v x_v^{\alpha_v}\right) \left(\sum_{\tau\in\mathcal{C}^{\text{max}}_G}(-1)^{|\mathbf{e}|}\mu_\tau(X_v+x_v,Y_e)\right),
\end{align}
where we used formula \eqref{eq:diffformsexpansion}, and where the differential $\dd$ in the definition of $\mu_\tau$ acts on the integration variables $x_v$ instead of the kinematic variables $X_v$. 

Formula \eqref{eq:FRWasform} shows that the cosmological wavefunction coefficients can be naturally written in terms of twisted integrals $\int u\,  \mu_\tau$ for $\tau\in\mathcal{C}^{\text{max}}_G$, where $u=\prod_v x_v^{\alpha_v}$. However, the differential equations that we will find in the next section require a slight modification of this set of functions. We first observe that there is yet another rewriting of the $\psi^{\text{FRW}}_G$
\begin{equation}\label{eq:FRWastubings}
\psi^{\text{FRW}}_G(X_v,Y_e)=\int \left(\prod_v x_v^{\alpha_v}\right) \left(\sum_{\tau\in\mathcal{C}_G}(-1)^{|\mathbf{e}|+|V_G|-n_\tau}\tilde\mu_\tau(X_v,Y_e)\right),
\end{equation}
where the sum runs over {\it all} $c$-tubings, $n_{\tau}$ is the number of tubes in tubing $\tau$, and the forms $\tilde\mu_\tau$ are defined as
\begin{equation}\label{eq:omegageneric}
\widetilde\mu_\tau=\dlog \frac{\widetilde{H}_{T_1}}{x_1}\wedge\ldots\wedge \dlog \frac{\widetilde{H}_{T_{|V_G|}}}{x_{|V_G|}}\,,
\end{equation} 
with
\begin{equation}
\widetilde{H}_T=H_T(X_v\to x_v+X_v)\,.
\end{equation}
The set of all $\tilde\mu_\tau$ for $\tau\in\mathcal{C}_G$ will form a set of logarithmic differential forms whose twisted integrals $\int u\, \widetilde{\mu}_\tau$ will form a closed set of differential equations, as we show in the next section.

\section{Differential Equations}\label{sec:diffeqs}
\subsection{Setup}
In this section we set the stage for our novel derivation of differential equations for cosmological wavefunction coefficients \eqref{eq:FRWintro}.
In our discussion, we will strongly rely on the fact that all functions of interest associated to a given graph $G$ can be written as twisted integrals
\begin{equation}\label{eq:twisted_integral}
\Psi=\int u\, \omega\,,
\end{equation}
where $\omega$ is a logarithmic differential form in the variables $x_v$, $v=1,\ldots,|V_G|$, and $u=\prod_{v}x_v^{\alpha_v}$ is the twist function.
For a given Feynman diagram $G$, the wavefunction coefficients are functions of variables $(X_v,Y_e)_{v\in V_G,e\in E_G}$, and therefore we aim at writing the differential equations in this kinematic space. For this purpose, we define the kinematic differential
\begin{equation}
\dd_{\text{kin}}=\sum_{v\in V_G}\dd X_v\frac{\partial}{\partial X_v}+\sum_{e\in E_G}\dd Y_e\frac{\partial}{\partial Y_e}\,.
\end{equation}
When acting with this kinematic differential on the cosmological wavefunction coefficients $\psi^{\text{FRW}}_{G}$, one usually generates new functions on the right hand side of the equations. However, for a given graph $G$, there is a finite set of functions that one can obtain -- the set of master integrals -- and therefore one can always close the set of differential equations. In the following, we will introduce a particular set of master integrals for a given graph, but in general we are interested in obtaining a set of equations
\begin{equation}\label{eq:diffeqsF}
\dd_{\text{kin}}\vec{F}=B\vec{F}\,,
\end{equation}
where $\vec{F}$ is a set of master integrals, and $B$ is a matrix with entries which are functions of the twist exponents $\alpha_i$ and (logarithmic) one-forms of the kinematic variables $(X_v,Y_e)$. Importantly, we want the matrix $B$ to be in the so-called {\it canonical form} (or $\epsilon$-form) \cite{Henn:2013pwa}, which additionally means that all its entries should vanish when $\alpha_v\to 0$, allowing for a perturbative solution to these equations as power series in the coefficients $\alpha_v$.

In our set of master integrals, each of them will be written as a twisted integral of the form \eqref{eq:twisted_integral}
\begin{equation}
\vec{F}=\int u\, \vec\omega\,,
\end{equation}
where $\vec{\omega}$ is a vector of differential forms, and therefore the set of differential equations \eqref{eq:diffeqsF} is equivalent, up to total derivative terms, to the following set of equations for $\vec\omega$
 \begin{equation}\label{eq:diffeqs}
 \dd_{\text{kin}}\vec{\omega}=B\wedge\vec{\omega}\,.
 \end{equation}

\subsection{The Power of Logarithmic Differential Forms}

In the following, we want to remain in the realm of logarithmic differential forms throughout all our calculations\footnote{See \cite{Gasparotto:2024bku} for a similar derivation.}. As we will see momentarily, this will immediately allow us to write equations \eqref{eq:diffeqsF} in the $\epsilon$-form. Let us take any function $\Psi$ that can be written as the twisted integral \eqref{eq:twisted_integral}. Acting with the kinematic differential $\dd_{\text{kin}}$ on $\Psi$ we get
\begin{align}\label{eq:dkinpsi}
\dd_{\text{kin}}\Psi=\dd_{\text{kin}}\int u\, \omega=\int (\dd_{\text{kin}}u\wedge \omega+u\, \dd_{\text{kin}}\omega)\,.
\end{align}
Since $\omega=\dlog \gamma_1\wedge\ldots\wedge\dlog \gamma_{|V_G|}$ is a logarithmic $|V_G|$-form on the space of $x_v$, then the action of the kinematic differential can be rewritten as
\begin{align}
\dd_{\text{kin}}\omega&=\dd_{\text{kin}}(\dlog\gamma_1\wedge \ldots\wedge\dlog\gamma_{|V_G|})\\
&=\sum_{i}(-1)^i(\dlog\gamma_1\wedge \ldots\wedge (\dd_{\text{kin}} \dlog\gamma_i)\wedge\ldots\wedge\dlog\gamma_{|V_G|})\\
&=-\sum_{i}(-1)^i(\dlog\gamma_1\wedge \ldots\wedge (\dd \,\dd_{\text{kin}}\hspace{-0.05cm} \log\gamma_i)\wedge\ldots\wedge\dlog\gamma_{|V_G|})\\
&=-\dd\sum_{i}(-1)^i(\dlog\gamma_1\wedge \ldots\wedge \dd_{\text{kin}} \hspace{-0.05cm}\log\gamma_i\wedge\ldots\wedge\dlog\gamma_{|V_G|})\\
&=:-\dd \tilde\omega\,,
\end{align}
where $\tilde\omega$ is a $(|V_G|-1)$-form of variables $x_v$ and a one-form of the kinematic variables $(X_v,Y_e)$.
Using this relation, and integrating by parts the second term in \eqref{eq:dkinpsi}, one gets
\begin{equation}
\dd_{\text{kin}}\Psi=\int (\dd_{\text{kin}}u\wedge \omega+\dd u\wedge \tilde\omega)=\int u\,(\dd_{\text{kin}}\log u\wedge \omega+\dlog u\wedge \tilde\omega)\,,
\end{equation}
where we dropped the boundary terms.
It is now very natural to recognise that the integrand in the formula above is a particular term, namely, the term of degree $|V_G|$ in the integrated variables and of degree one in kinematic variables, of the following differential form
\begin{equation}
\dd_{\text{kin}}\log u\wedge \omega+\dlog u\wedge \tilde\omega=\Dlog u\wedge \Dlog\gamma_1\wedge\ldots\wedge \Dlog\gamma_{|V_G|}\Big|^{(|V_G|,1)}=:\Dlog u\wedge\Omega\Big|^{(|V_G|,1)}\,,
\end{equation}
where $\mathfrak{D}=\dd_{\text{kin}}+\dd$ and $\Omega$ is the differential form $\omega$ with the differential $\dd$ replaced by $\mathfrak{D}$. Since the integration, that is done over the variables $x_v$, $v=1,\ldots,|V_G|$, selects the term of $\Omega$ of degree $|V_G|$ in the integrated variables, the differential equation finally reduces to the following 
\begin{equation}\label{eq:diffeq2}
\dd_{\text{kin}}\Psi=\int u \,(\Dlog u\wedge\Omega)\,.
\end{equation}
Notice that on the right hand side of this equation, there are many terms that integrate to zero because of the degree of the form. Therefore, in the procedure explained in this section, we have added artificial zeros that allow for a more symmetric version of the integrand, where the kinematic differentials and the differentials for the integrated variables are treated on the same footing. Since the right-hand side is again a twisted integral with the integrand $\Dlog u\wedge\Omega$, in order to derive our differential equations for $\Psi$, it is sufficient to find the action of $\Dlog u$ on $\Omega$ and then decompose it in our set of differential forms leading to
\begin{equation}\label{eq:diffeq3}
\dd_{\text{kin}}\Psi= \int u\left(\sum \Dlog \kappa_i(X_v,Y_e) \wedge \omega_i\right) =\sum_i  \Dlog \kappa_i(X_v,Y_e) F_i\,,
\end{equation}
where, importantly, $\kappa_i(X_v,Y_e)$ must be independent of the integration variables $x_v$. In the following, we will show that such decomposition is always possible when acting on any of the  differential forms $\widetilde\mu_\tau$ in \eqref{eq:omegageneric}.
Crucially, since all terms on the right hand side of \eqref{eq:diffeq2} contain $\Dlog u$, then each term is proportional to one of the twist parameters $\alpha_v$, and therefore the resulting differential equations are guaranteed to be in the canonical form!

\subsection{Useful Relations}
As we mentioned in the previous section, in order to derive the set of equations for $\Psi$, we need to be able to decompose $\Dlog u\wedge\Omega$ as a sum of forms $\sum_i \Dlog \kappa_i\wedge \omega_i$ such that $\kappa_i$ is independent of the integration variables $x_v$ and $\omega_i$ are elements of our set of differential forms. In this section we demonstrate how this works for any element of our set $\widetilde\mu_\tau\big|_{\dd\to\mathfrak{D}}$ for any $c$-tubing $\tau\in\mathcal{C}_G$.   First, we notice that we can expand the canonical form $\widetilde\mu_\tau\big|_{\dd\to\mathfrak{D}}$ as
\begin{align}\label{eq:mutildeexp}
\widetilde\mu_\tau\big|_{\dd\to\mathfrak{D}}&= \Dlog \widetilde{H}_{T_1}\wedge\Dlog \widetilde{H}_{T_2}\wedge\ldots\wedge \Dlog \widetilde{H}_{T_{|V_G|}}\nonumber\\
&-\Dlog x_{1}\wedge\Dlog \widetilde{H}_{T_2}\wedge\ldots\wedge \Dlog \widetilde{H}_{T_{|V_G|}}+\ldots\,,
\end{align}
where each term in the expansion has either $x_v$ or $\widetilde{H}_{T_v}$ at position $v$ of the wedge product.
Moreover, we emphasize that
\begin{equation}
\widetilde{H}_{T}=\sum_{v\in T}x_v+\ldots\,,
\end{equation}
where the ellipsis contains only kinematic variables. 
The action of $\Dlog u=\sum_v \alpha_v \Dlog x_v$ on $\widetilde\mu_\tau\big|_{\dd\to\mathfrak{D}}$ is therefore of the form
\begin{equation}\label{eq:exptemp}
\Dlog x_i\wedge\Dlog \overline{H}_{T_1}\wedge\Dlog \overline{H}_{T_2}\wedge\ldots\wedge \Dlog \overline{H}_{T_{|V_G|}}\,,
\end{equation}
where either $\overline{H}_{T_v}=\widetilde{H}_{T_v}$ or $\overline{H}_{T_v}=x_v$, and all tubes appearing in \eqref{eq:exptemp} are compatible as $b$-tubes. Importantly, when looking at these expressions on a case-by-case basis, it is always possible to find a linear combination of the arguments of $\Dlog$'s in \eqref{eq:exptemp} of the form
\begin{equation}
\pm\overline{H}_{T_{i_1}}\pm\ldots\pm \overline{H}_{T_{i_p}}-x_i\,,
\end{equation}
 that is independent of the integration variables $x_v$. Notice that the coefficients in this linear combination are either $+1$ or $-1$.
Then, using very simple relations that hold true for logarithmic differential forms
\begin{equation}\label{eq:dlogrelation}
\Dlog \gamma_1\wedge \ldots \wedge\Dlog \gamma_p = \Dlog (\pm \gamma_1\pm \ldots\pm\gamma_p)\wedge\Dlog \frac{\gamma_2}{\gamma_1}\wedge\ldots\wedge \Dlog \frac{\gamma_p}{\gamma_1}\,,
\end{equation}
it is possible to eliminate all integration variables from one of the $\Dlog$'s. After applying this relation to each term in \eqref{eq:mutildeexp}, they can be written as
\begin{equation}
\Dlog \kappa(X_v,Y_e)\wedge\Dlog \frac{\overline{H}_{T_{i_1}}}{x_i}\wedge\ldots\wedge\Dlog \frac{\overline{H}_{T_{i_p}}}{x_i}\wedge\Dlog \overline{H}_{T_{i_{p+1}}}\wedge\ldots\wedge \Dlog \overline{H}_{T_{i_{|V_G|}}}\,,
\end{equation}
where the first factor does not depend on the integration variables, and the remaining $|V_G|$-form can be decomposed as a sum of the canonical forms $\widetilde\mu_\tau$, leading to \eqref{eq:diffeq3}. Surprisingly, in all examples that we have studied, the function $\kappa(X_v,Y_e)$ is always the region variable $R_r$ for one of the regions of graph $G$, $r\in\mathcal{R}_{G}$. We demostrate that this is indeed the case in the examples in the following section, and we conjecture that it is true for all graphs $G$; however, a proof of this statement is beyond the scope of this paper.

\subsection{Explicit Form of Differential Equations}
We are finally ready to provide an explicit form of differential equations for cosmological wavefunction coefficients based on the approach explained above. For a given graph $G$, we define the set of functions
\begin{equation}
F_\tau=\int u \, \widetilde\mu_\tau\,,
\end{equation}
where $\tau\in\mathcal{C}_G$ is a $c$-tubing on $G$, and $\widetilde\mu_\tau$ is defined in \eqref{eq:omegageneric}. The cosmological wavefunction coefficient $\psi_G^{\text{FRW}}$ can be decomposed in terms of these functions as
\begin{equation}
\psi_G^{\text{FRW}}=\sum_{\tau\in\mathcal{C}_G}(-1)^{|\mathbf{e}|+|V_G|-n_\tau}F_\tau\,.
\end{equation}
Using the results of this section, the action of $\Dlog u\wedge$ on the canonical forms $\widetilde\mu_\tau$ can be calculated to be
\begin{equation}\label{eq:actionDlog}
\Dlog x_v\wedge \widetilde\mu_\tau= \sum_{\tau'\in\mathcal{R}_{\tau,v}}\Dlog R_{\tau',v}\wedge\left(\sum_{\tau''\in \mathcal{T}_{\tau',v}}(-1)^{n_{\tau'}-n_{\tau''}}\,\widetilde\mu_{\tau''}\right),
\end{equation}
where $\mathcal{R}_{\tau,v}$ is the set of $c$-tubings obtained recursively by removing from the tubing $\tau$ any child tube of the parent tube $q_\tau(v)$, and by repeating this operation on each tubing found this way until no more tubes can be removed. Then, for any $\tau'\in \mathcal{R}_{\tau,v}$, the variable $R_{\tau',v}$ is the region variable associated to the region whose parent tube is $q_{\tau'}(v)$. Finally, $\mathcal{T}_{\tau',v}$ is the set of $c$-tubings obtained from $\tau'$ by removing any subset of children tubes of the region whose parent tube is $q_{\tau'}(v)$.

Then the formula \eqref{eq:actionDlog} implies the following set of differential equations for $F_\tau$
\begin{equation}\label{eq:finaleqs}
\dd_{\text{kin}}F_\tau=\sum_{v\in V_G}\alpha_v\left( \sum_{\tau'\in\mathcal{R}_{\tau,v}}\Dlog R_{\tau',v}\left(\sum_{\tau''\in \mathcal{T}_{\tau',v}}(-1)^{n_{\tau'}-n_{\tau''}}\,F_{\tau''}\right)\right).
\end{equation}
%
By introducing the constant matrices $C^{(v)}_{\tau'}$ with entries $(C^{(v)}_{\tau'})_{\tau\tau''}=(-1)^{n_{\tau'}-n_{\tau''}}$ when $\tau'\in \mathcal{R}_{\tau,v}$ and $\tau''\in\mathcal{T}_{\tau',v}$, and $(C^{(v)}_{\tau'})_{\tau\tau''}=0$ otherwise, these equations can be recast into a vector form
\begin{equation}\label{eq:canonicaldiffeq}
\dd_{\text{kin}}\vec{F}=\sum_{v\in V_G}\alpha_v\left[\left(\sum_{\tau'\in\mathcal{R}_{v}}C^{(v)}_{\tau'} \Dlog R_{\tau',v} \right)\cdot \vec{{F}}\right],
\end{equation}
where $\mathcal{R}_{v}$ is the union of sets $\mathcal{R}_{\tau,v}$ where $\tau$ ranges over all the $c$-tubings of the basis. 
We have therefore arrived to a closed set of $|\mathcal{C}_G|$ differential equations in $|V_G|+|E_G|$ variables that are canonical in each twist parameter, with logarithmic integration kernels given by region variables. The general solution of these equations can be written in terms of a path-ordered exponential that can be evaluated perturbatively in the twist parameters. The class of special functions one expects to find is that of \textit{multiple polylogarithms}, iterated integrals whose set of singularities is given by the \textit{alphabet} formed by all the region variables in \eqref{eq:canonicaldiffeq}.

\subsection{Diagonalisation of Differential Equations Coefficients}
As a final comment regarding the resulting equations, we provide an interesting observation regarding the matrix of coefficients in equation \eqref{eq:finaleqs}. We have found that it can be easily diagonalised, and the resulting eigensystem can be again interpreted in terms of the same notion of tubings on graph $G$. After checking  various examples for basic graphs, we conjecture that the eigenvectors of the coefficient matrix for the differential equations \eqref{eq:finaleqs} are given by 
\begin{equation}\label{eq:eigenvector}
\overline{F}_{\tau}=\sum_{\tau'\subset \tau}\left(\prod_{T\in\tau'}\left(\frac{\left(\sum_{v\in r_{\tau,T}^{\text{in}}}\alpha_{v}\right)R_{r_{\tau,T}^{\text{out}}}}{\left(\sum_{v\in r_{\tau,T}^{\text{out}}}\alpha_{v}\right)R_{r_{\tau,T}^{\text{in}}}}-1\right)\right)F_{\tau'}\,,
\end{equation}
where the sum runs over all $c$-subtubings $\tau'$, i.e.~over the set of $c$-tubings that can be obtained from $\tau$ by removing any number of tubes, and the regions $r_{\tau,T}^{\text{out}}$ and $r_{\tau,T}^{\text{in}}$ are defined in section \eqref{sec:tubings}.
After writing $\overline{F}_\tau=\int u\, \overline\mu_\tau$ as a twisted integral, the eigenvalues corresponding to the eigenvectors in \eqref{eq:eigenvector} can be obtained by computing the action of $\Dlog u$ on $\overline{\mu}_\tau$, which reads
\begin{equation}\label{eq:eigenvalue}
\Dlog u\wedge \overline{\mu}_\tau=\left(\sum_{v}\alpha_v \Dlog R_{r_v}\right)\, \overline{\mu}_\tau\,,
\end{equation}
where the coefficient of the eigenvalues in front of $\alpha_v$ is just the region variable for the region $r_v$ containing vertex $v$. We conjecture that formulae \eqref{eq:eigenvector} and \eqref{eq:eigenvalue} give eigenvectors and eigenvalues of the matrix of coefficients in the differential equations \eqref{eq:finaleqs} for any graph $G$. We have checked it for a large number of graphs, tree and with cycles, and leave a general proof of this statement for future work.
\section{Examples}\label{sec:examples}
In this section we provide explicit examples of all notions introduced in the previous sections, and provide an explicit form of the differential equations obtained using the method explained in section \ref{sec:diffeqs}.

\newpage
\subsection{Tree Graphs}

{\bf $\bullet$ Path graph \begin{tikzpicture}[scale=0.6]
    \coordinate (A) at (0,0);
    \coordinate (B) at (1/2,0);
    \draw[thick] (A) -- (B);
    \fill[black] (A) circle (2pt);
    \fill[black] (B) circle (2pt);
\end{tikzpicture}}

We start by considering the simplest non-trivial case, that is the path graph with two vertices
\begin{equation}
\begin{tikzpicture}[scale=1.2]
    \coordinate (A) at (0,0);
    \coordinate (B) at (1,0);
    \coordinate (C) at (-1,0);
    \draw[thick] (A) -- (B);
    \fill[black] (A) circle (2pt) node[above] {$X_1$};
    \fill[black] (B) circle (2pt) node[above] {$X_2$};
    \node at ($(A)!0.5!(B)$) [below] {$Y_{1,2}$};
    \node at (C) {$G=$};
\end{tikzpicture}
\end{equation}
For tree graphs, the notions of $b$-tubes and $u$-tubes are identical, and for the path graph with two vertices we have three tubes:
\begin{equation}
B_{\begin{tikzpicture}[scale=0.6]
    \coordinate (A) at (0,0);
    \coordinate (B) at (1/2,0);
    \draw[thick] (A) -- (B);
    \fill[black] (A) circle (2pt);
    \fill[black] (B) circle (2pt);
\end{tikzpicture}}
=U_{\begin{tikzpicture}[scale=0.6]
    \coordinate (A) at (0,0);
    \coordinate (B) at (1/2,0);
    \draw[thick] (A) -- (B);
    \fill[black] (A) circle (2pt);
    \fill[black] (B) circle (2pt);
\end{tikzpicture}}
=\{
\raisebox{-0.15cm}{\input{Figures/graph_path_2_u_tubes.tex}}
\}\,.
\end{equation}
However, the sets of $u$-tubings and $b$-tubings are different, with every $u$-tubing being a $b$-tubing, but not vice versa. In this case the set of all $u$-tubings is
\begin{equation}
\mathcal{U}_{\begin{tikzpicture}[scale=0.6]
    \coordinate (A) at (0,0);
    \coordinate (B) at (1/2,0);
    \draw[thick] (A) -- (B);
    \fill[black] (A) circle (2pt);
    \fill[black] (B) circle (2pt);
\end{tikzpicture}}=
\{
\raisebox{-0.15cm}{\input{Figures/graph_path_2_u_tubings.tex}}
\}\,,
\end{equation}
while there are eight $b$-tubings composed of any subset of the set $B_{\begin{tikzpicture}[scale=0.6]
    \coordinate (A) at (0,0);
    \coordinate (B) at (1/2,0);
    \draw[thick] (A) -- (B);
    \fill[black] (A) circle (2pt);
    \fill[black] (B) circle (2pt);
\end{tikzpicture}}$.
To construct the flat-space wavefunction we will only need the set containing maximal $b$-tubings, which in this case is
\begin{equation}
\mathcal{B}^{\text{max}}_{\begin{tikzpicture}[scale=0.6]
    \coordinate (A) at (0,0);
    \coordinate (B) at (1/2,0);
    \draw[thick] (A) -- (B);
    \fill[black] (A) circle (2pt);
    \fill[black] (B) circle (2pt);
\end{tikzpicture}}=
\{
\raisebox{-0.15cm}{\input{Figures/graph_path_2_b_tubings.tex}}
\}\,.
\end{equation}
Using formulae \eqref{eq:wavefunction} and \eqref{eq:amplitube} we find
\begin{align}
\psi^{\text{flat}}_{\begin{tikzpicture}[scale=0.6]
    \coordinate (A) at (0,0);
    \coordinate (B) at (1/2,0);
    \draw[thick] (A) -- (B);
    \fill[black] (A) circle (2pt);
    \fill[black] (B) circle (2pt);
\end{tikzpicture}}
&=\frac{2Y_{1,2}}{
H_{\begin{tikzpicture}[scale=0.6]
    \coordinate (A1) at (0,0);
    \coordinate (B1) at (1,0);
    \draw[thick] (A1) -- (B1);
        \draw[thick, black] (1/2,0) ellipse (0.75cm and 0.3cm);
    \fill[black] (A1) circle (2pt);
    \fill[black] (B1) circle (2pt);
    \end{tikzpicture}}\,
H_{\begin{tikzpicture}[scale=0.6]
    \coordinate (A2) at (0,0);
    \coordinate (B2) at (1,0);
    \draw[thick] (A2) -- (B2);
    \draw[thick, black] (0,0) ellipse (0.15cm and 0.15cm);
    \fill[black] (A2) circle (2pt);
    \fill[black] (B2) circle (2pt);
\end{tikzpicture}}\,
H_{\begin{tikzpicture}[scale=0.6]
    \coordinate (A3) at (4,0);
    \coordinate (B3) at (5,0);
    \draw[thick] (A3) -- (B3);
    \draw[thick, black] (5,0) ellipse (0.15cm and 0.15cm);
    \fill[black] (A3) circle (2pt);
    \fill[black] (B3) circle (2pt);
\end{tikzpicture}}}
=\frac{2Y_{1,2}}{(X_1+X_2)(X_1+Y_{1,2})(X_2+Y_{1,2})}\\
&=\left(\frac{1}{(X_1+X_2)(X_1+Y_{1,2})}+\frac{1}{(X_1+X_2)(X_2+Y_{1,2})}\right)-\frac{1}{(X_1+Y_{1,2})(X_2+Y_{1,2})}\\
&=A_{\begin{tikzpicture}[scale=0.6]
    \coordinate (A) at (0,0);
    \coordinate (B) at (1/2,0);
    \draw[thick] (A) -- (B);
    \fill[black] (A) circle (2pt);
    \fill[black] (B) circle (2pt);
\end{tikzpicture}}-A_{\begin{tikzpicture}[scale=0.6]
    \coordinate (A) at (0,0);
    \coordinate (B) at (1/2,0);
    \fill[black] (A) circle (2pt);
    \fill[black] (B) circle (2pt);
\end{tikzpicture}}
\end{align}
When uplifted to differential forms, this relation is
\begin{align}
\widetilde{\Omega}^{\text{flat}}_{\begin{tikzpicture}[scale=0.6]
    \coordinate (A) at (0,0);
    \coordinate (B) at (1/2,0);
    \draw[thick] (A) -- (B);
    \fill[black] (A) circle (2pt);
    \fill[black] (B) circle (2pt);
\end{tikzpicture}}=
\omega_{\begin{tikzpicture}[scale=0.6]
    \coordinate (A) at (0,0);
    \coordinate (B) at (1/2,0);
    \draw[thick] (A) -- (B);
    \fill[black] (A) circle (2pt);
    \fill[black] (B) circle (2pt);
\end{tikzpicture}}
-\omega_{\begin{tikzpicture}[scale=0.6]
    \coordinate (A) at (0,0);
    \coordinate (B) at (1/2,0);
    \fill[black] (A) circle (2pt);
    \fill[black] (B) circle (2pt);
\end{tikzpicture}}
&=\left(\dlog H_{\begin{tikzpicture}[scale=0.6]
    \coordinate (A2) at (0,0);
    \coordinate (B2) at (1,0);
    \draw[thick] (A2) -- (B2);
    \draw[thick, black] (0,0) ellipse (0.15cm and 0.15cm);
    \fill[black] (A2) circle (2pt);
    \fill[black] (B2) circle (2pt);
\end{tikzpicture}}
\wedge \dlog H_{\begin{tikzpicture}[scale=0.6]
    \coordinate (A1) at (0,0);
    \coordinate (B1) at (1,0);
    \draw[thick] (A1) -- (B1);
        \draw[thick, black] (1/2,0) ellipse (0.75cm and 0.3cm);
    \fill[black] (A1) circle (2pt);
    \fill[black] (B1) circle (2pt);
    \end{tikzpicture}}
+\dlog H_{\begin{tikzpicture}[scale=0.6]
    \coordinate (A1) at (0,0);
    \coordinate (B1) at (1,0);
    \draw[thick] (A1) -- (B1);
        \draw[thick, black] (1/2,0) ellipse (0.75cm and 0.3cm);
    \fill[black] (A1) circle (2pt);
    \fill[black] (B1) circle (2pt);
    \end{tikzpicture}}
\wedge \dlog H_{\begin{tikzpicture}[scale=0.6]
    \coordinate (A3) at (4,0);
    \coordinate (B3) at (5,0);
    \draw[thick] (A3) -- (B3);
    \draw[thick, black] (5,0) ellipse (0.15cm and 0.15cm);
    \fill[black] (A3) circle (2pt);
    \fill[black] (B3) circle (2pt);
\end{tikzpicture}}\right)\nonumber\\
&-\dlog H_{\begin{tikzpicture}[scale=0.6]
    \coordinate (A2) at (0,0);
    \coordinate (B2) at (1,0);
    \draw[thick] (A2) -- (B2);
    \draw[thick, black] (0,0) ellipse (0.15cm and 0.15cm);
    \fill[black] (A2) circle (2pt);
    \fill[black] (B2) circle (2pt);
\end{tikzpicture}}
\wedge \dlog H_{\begin{tikzpicture}[scale=0.6]
    \coordinate (A3) at (4,0);
    \coordinate (B3) at (5,0);
    \draw[thick] (A3) -- (B3);
    \draw[thick, black] (5,0) ellipse (0.15cm and 0.15cm);
    \fill[black] (A3) circle (2pt);
    \fill[black] (B3) circle (2pt);
\end{tikzpicture}}\\
&=-\dlog \frac{X_1+Y_{1,2}}{X_1+X_2}\wedge \dlog \frac{X_2+Y_{1,2}}{X_1+X_2}\,.
\end{align}
The relation between the canonical forms $\widetilde{\Omega}$ and $\omega$ in the formula above can be understood geometrically as in Fig.~\ref{fig:p2geom}.
\begin{figure}[t]
\input{Figures/graph_path_2_regions.tex}
\caption{Geometric interpretation of the relation between $\widetilde{\Omega}$ and $\omega$ for the path graph with two vertices.}
\label{fig:p2geom}
\end{figure}

The set of differential forms for this graph is
\begin{equation}
v_{\begin{tikzpicture}[scale=0.6]
    \coordinate (A) at (0,0);
    \coordinate (B) at (1/2,0);
    \draw[thick] (A) -- (B);
    \fill[black] (A) circle (2pt);
    \fill[black] (B) circle (2pt);
\end{tikzpicture}}=\left\{\widetilde{\mu}_{\begin{tikzpicture}[scale=0.6]
    \coordinate (A1) at (0,0);
    \coordinate (B1) at (1,0);
    \draw[thick] (A1) -- (B1);
        \draw[thick, red] (1/2,0) ellipse (0.75cm and 0.3cm);
    \fill[black] (A1) circle (2pt);
    \fill[black] (B1) circle (2pt);
\end{tikzpicture}},\widetilde{\mu}_{\begin{tikzpicture}[scale=0.6]
    \coordinate (A1) at (0,0);
    \coordinate (B1) at (1,0);
    \draw[thick] (A1) -- (B1);
        \draw[thick, red] (1/2,0) ellipse (0.75cm and 0.3cm);
            \draw[thick, black] (0,0) ellipse (0.15cm and 0.15cm);
    \fill[black] (A1) circle (2pt);
    \fill[black] (B1) circle (2pt);
\end{tikzpicture}},\widetilde{\mu}_{\begin{tikzpicture}[scale=0.6]
    \coordinate (A1) at (0,0);
    \coordinate (B1) at (1,0);
    \draw[thick] (A1) -- (B1);
        \draw[thick, red] (1/2,0) ellipse (0.75cm and 0.3cm);
            \draw[thick, black] (1,0) ellipse (0.15cm and 0.15cm);
    \fill[black] (A1) circle (2pt);
    \fill[black] (B1) circle (2pt);
\end{tikzpicture}},\widetilde{\mu}_{\begin{tikzpicture}[scale=0.6]
    \coordinate (A1) at (0,0);
    \coordinate (B1) at (1,0);
    \draw[thick] (A1) -- (B1);
            \draw[thick, red] (0,0) ellipse (0.15cm and 0.15cm);
            \draw[thick, red] (1,0) ellipse (0.15cm and 0.15cm);
    \fill[black] (A1) circle (2pt);
    \fill[black] (B1) circle (2pt);
\end{tikzpicture}}\right\}\,,
\end{equation}
where the forms have the following explicit expressions
\begin{align}
\widetilde{\mu}_{\begin{tikzpicture}[scale=0.6]
    \coordinate (A1) at (0,0);
    \coordinate (B1) at (1,0);
    \draw[thick] (A1) -- (B1);
        \draw[thick, red] (1/2,0) ellipse (0.75cm and 0.3cm);
    \fill[black] (A1) circle (2pt);
    \fill[black] (B1) circle (2pt);
\end{tikzpicture}}&=\Dlog \frac{x_1+X_1+x_2+X_2}{x_1}\wedge \Dlog \frac{x_1+X_1+x_2+X_2}{x_2}\,,\\
\widetilde{\mu}_{\begin{tikzpicture}[scale=0.6]
    \coordinate (A1) at (0,0);
    \coordinate (B1) at (1,0);
    \draw[thick] (A1) -- (B1);
        \draw[thick, red] (1/2,0) ellipse (0.75cm and 0.3cm);
            \draw[thick, black] (0,0) ellipse (0.15cm and 0.15cm);
    \fill[black] (A1) circle (2pt);
    \fill[black] (B1) circle (2pt);
\end{tikzpicture}}&=\Dlog \frac{x_1+X_1+Y_{1,2}}{x_1}\wedge \Dlog \frac{x_1+X_1+x_2+X_2}{x_2}\,,\\
\widetilde{\mu}_{\begin{tikzpicture}[scale=0.6]
    \coordinate (A1) at (0,0);
    \coordinate (B1) at (1,0);
    \draw[thick] (A1) -- (B1);
        \draw[thick, red] (1/2,0) ellipse (0.75cm and 0.3cm);
            \draw[thick, black] (1,0) ellipse (0.15cm and 0.15cm);
    \fill[black] (A1) circle (2pt);
    \fill[black] (B1) circle (2pt);
\end{tikzpicture}}&=\Dlog \frac{x_1+X_1+x_2+X_2}{x_1}\wedge \Dlog \frac{x_2+X_2+Y_{1,2}}{x_2}\,,\\
\widetilde{\mu}_{\begin{tikzpicture}[scale=0.6]
    \coordinate (A1) at (0,0);
    \coordinate (B1) at (1,0);
    \draw[thick] (A1) -- (B1);
            \draw[thick, red] (0,0) ellipse (0.15cm and 0.15cm);
            \draw[thick, red] (1,0) ellipse (0.15cm and 0.15cm);
    \fill[black] (A1) circle (2pt);
    \fill[black] (B1) circle (2pt);
\end{tikzpicture}}&=\Dlog \frac{x_1+X_1+Y_{1,2}}{x_1}\wedge \Dlog \frac{x_2+X_2+Y_{1,2}}{x_2}\,.
\end{align}
Notice that all the logarithmic differential forms $\widetilde{\mu}$ are associated to \emph{compact} regions. 
Then, the flat-space canonical form $\widetilde{\Omega}^{\text{flat}}_{\begin{tikzpicture}[scale=0.6]
    \coordinate (A) at (0,0);
    \coordinate (B) at (1/2,0);
    \draw[thick] (A) -- (B);
    \fill[black] (A) circle (2pt);
    \fill[black] (B) circle (2pt);
\end{tikzpicture}}$ can be written as
\begin{equation}
\widetilde{\Omega}^{\text{flat}}_{\begin{tikzpicture}[scale=0.6]
    \coordinate (A) at (0,0);
    \coordinate (B) at (1/2,0);
    \draw[thick] (A) -- (B);
    \fill[black] (A) circle (2pt);
    \fill[black] (B) circle (2pt);
\end{tikzpicture}}(X_v+x_v,Y_e)=\left(
\widetilde{\mu}_{\begin{tikzpicture}[scale=0.6]
    \coordinate (A1) at (0,0);
    \coordinate (B1) at (1,0);
    \draw[thick] (A1) -- (B1);
        \draw[thick, red] (1/2,0) ellipse (0.75cm and 0.3cm);
            \draw[thick, black] (0,0) ellipse (0.15cm and 0.15cm);
    \fill[black] (A1) circle (2pt);
    \fill[black] (B1) circle (2pt);
\end{tikzpicture}}+\widetilde{\mu}_{\begin{tikzpicture}[scale=0.6]
    \coordinate (A1) at (0,0);
    \coordinate (B1) at (1,0);
    \draw[thick] (A1) -- (B1);
        \draw[thick, red] (1/2,0) ellipse (0.75cm and 0.3cm);
            \draw[thick, black] (1,0) ellipse (0.15cm and 0.15cm);
    \fill[black] (A1) circle (2pt);
    \fill[black] (B1) circle (2pt);
\end{tikzpicture}}-
\widetilde{\mu}_{\begin{tikzpicture}[scale=0.6]
    \coordinate (A1) at (0,0);
    \coordinate (B1) at (1,0);
    \draw[thick] (A1) -- (B1);
        \draw[thick, red] (1/2,0) ellipse (0.75cm and 0.3cm);
    \fill[black] (A1) circle (2pt);
    \fill[black] (B1) circle (2pt);
\end{tikzpicture}}\right)
-\widetilde{\mu}_{\begin{tikzpicture}[scale=0.6]
    \coordinate (A1) at (0,0);
    \coordinate (B1) at (1,0);
    \draw[thick] (A1) -- (B1);
            \draw[thick, red] (0,0) ellipse (0.15cm and 0.15cm);
            \draw[thick, red] (1,0) ellipse (0.15cm and 0.15cm);
    \fill[black] (A1) circle (2pt);
    \fill[black] (B1) circle (2pt);
\end{tikzpicture}}\,.
\end{equation}
The action of the $\Dlog u_{\begin{tikzpicture}[scale=0.6]
    \coordinate (A) at (0,0);
    \coordinate (B) at (1/2,0);
    \draw[thick] (A) -- (B);
    \fill[black] (A) circle (2pt);
    \fill[black] (B) circle (2pt);
\end{tikzpicture}}=\alpha_1 \Dlog x_1+\alpha_2 \Dlog x_2$
on each of the forms $\widetilde\mu_\tau$ can be easily derived and reads:
\begin{align}
\Dlog u_{\begin{tikzpicture}[scale=0.6]
    \coordinate (A) at (0,0);
    \coordinate (B) at (1/2,0);
    \draw[thick] (A) -- (B);
    \fill[black] (A) circle (2pt);
    \fill[black] (B) circle (2pt);
\end{tikzpicture}} \wedge \widetilde{\mu}_{\begin{tikzpicture}[scale=0.6]
    \coordinate (A1) at (0,0);
    \coordinate (B1) at (1,0);
    \draw[thick] (A1) -- (B1);
        \draw[thick, red] (1/2,0) ellipse (0.75cm and 0.3cm);
    \fill[black] (A1) circle (2pt);
    \fill[black] (B1) circle (2pt);
\end{tikzpicture}}&= 
(\alpha_1+\alpha_2)\Dlog R_{\begin{tikzpicture}[scale=0.6]
    \coordinate (A1) at (0,0);
    \coordinate (B1) at (1,0);
            \draw[thick, red, fill=gray, opacity=0.5] (1/2,0) ellipse (0.75cm and 0.3cm);
    \draw[thick] (A1) -- (B1);
        \draw[thick, red] (1/2,0) ellipse (0.75cm and 0.3cm);
    \fill[black] (A1) circle (2pt);
    \fill[black] (B1) circle (2pt);
\end{tikzpicture}}
\wedge \widetilde{\mu}_{\begin{tikzpicture}[scale=0.6]
    \coordinate (A1) at (0,0);
    \coordinate (B1) at (1,0);
    \draw[thick] (A1) -- (B1);
        \draw[thick, red] (1/2,0) ellipse (0.75cm and 0.3cm);
    \fill[black] (A1) circle (2pt);
    \fill[black] (B1) circle (2pt);
\end{tikzpicture}}\,,\\
\Dlog u_{\begin{tikzpicture}[scale=0.6]
    \coordinate (A) at (0,0);
    \coordinate (B) at (1/2,0);
    \draw[thick] (A) -- (B);
    \fill[black] (A) circle (2pt);
    \fill[black] (B) circle (2pt);
\end{tikzpicture}}\wedge \widetilde{\mu}_{\begin{tikzpicture}[scale=0.6]
    \coordinate (A1) at (0,0);
    \coordinate (B1) at (1,0);
    \draw[thick] (A1) -- (B1);
        \draw[thick, red] (1/2,0) ellipse (0.75cm and 0.3cm);
            \draw[thick, black] (0,0) ellipse (0.15cm and 0.15cm);
    \fill[black] (A1) circle (2pt);
    \fill[black] (B1) circle (2pt);
\end{tikzpicture}}&= \alpha_1\,
\Dlog R_{\begin{tikzpicture}[scale=0.6]
    \coordinate (A1) at (0,0);
    \coordinate (B1) at (1,0);
    \draw[thick] (A1) -- (B1);

    \draw[thick, black, fill=gray,opacity=0.5] (0,0) ellipse (0.15cm and 0.15cm);
            \draw[thick, red] (1/2,0) ellipse (0.75cm and 0.3cm);
    \fill[black] (A1) circle (2pt);
    \fill[black] (B1) circle (2pt);
\end{tikzpicture}}
\wedge \widetilde{\mu}_{\begin{tikzpicture}[scale=0.6]
    \coordinate (A1) at (0,0);
    \coordinate (B1) at (1,0);
    \draw[thick] (A1) -- (B1);
        \draw[thick, red] (1/2,0) ellipse (0.75cm and 0.3cm);
            \draw[thick, black] (0,0) ellipse (0.15cm and 0.15cm);
    \fill[black] (A1) circle (2pt);
    \fill[black] (B1) circle (2pt);
\end{tikzpicture}}\nonumber\\
&+\alpha_2\left(
\Dlog R_{\input{Figures/graph_path_2_region_12_1.tex}}
\wedge \widetilde{\mu}_{\begin{tikzpicture}[scale=0.6]
    \coordinate (A1) at (0,0);
    \coordinate (B1) at (1,0);
    \draw[thick] (A1) -- (B1);
        \draw[thick, red] (1/2,0) ellipse (0.75cm and 0.3cm);
            \draw[thick, black] (0,0) ellipse (0.15cm and 0.15cm);
    \fill[black] (A1) circle (2pt);
    \fill[black] (B1) circle (2pt);
\end{tikzpicture}}
+\Dlog \frac{R_{\begin{tikzpicture}[scale=0.6]
    \coordinate (A1) at (0,0);
    \coordinate (B1) at (1,0);
            \draw[thick, red, fill=gray, opacity=0.5] (1/2,0) ellipse (0.75cm and 0.3cm);
    \draw[thick] (A1) -- (B1);
        \draw[thick, red] (1/2,0) ellipse (0.75cm and 0.3cm);
    \fill[black] (A1) circle (2pt);
    \fill[black] (B1) circle (2pt);
\end{tikzpicture}}}{R_{\input{Figures/graph_path_2_region_12_1.tex}}}
\wedge \widetilde{\mu}_{\begin{tikzpicture}[scale=0.6]
    \coordinate (A1) at (0,0);
    \coordinate (B1) at (1,0);
    \draw[thick] (A1) -- (B1);
        \draw[thick, red] (1/2,0) ellipse (0.75cm and 0.3cm);
    \fill[black] (A1) circle (2pt);
    \fill[black] (B1) circle (2pt);
\end{tikzpicture}}\right),\\
\Dlog u_{\begin{tikzpicture}[scale=0.6]
    \coordinate (A) at (0,0);
    \coordinate (B) at (1/2,0);
    \draw[thick] (A) -- (B);
    \fill[black] (A) circle (2pt);
    \fill[black] (B) circle (2pt);
\end{tikzpicture}} \wedge \widetilde{\mu}_{\begin{tikzpicture}[scale=0.6]
    \coordinate (A1) at (0,0);
    \coordinate (B1) at (1,0);
    \draw[thick] (A1) -- (B1);
        \draw[thick, red] (1/2,0) ellipse (0.75cm and 0.3cm);
            \draw[thick, black] (1,0) ellipse (0.15cm and 0.15cm);
    \fill[black] (A1) circle (2pt);
    \fill[black] (B1) circle (2pt);
\end{tikzpicture}}&= \alpha_1\left(
\Dlog R_{\input{Figures/graph_path_2_region_12_2.tex}}
\wedge \widetilde{\mu}_{\begin{tikzpicture}[scale=0.6]
    \coordinate (A1) at (0,0);
    \coordinate (B1) at (1,0);
    \draw[thick] (A1) -- (B1);
        \draw[thick, red] (1/2,0) ellipse (0.75cm and 0.3cm);
            \draw[thick, black] (1,0) ellipse (0.15cm and 0.15cm);
    \fill[black] (A1) circle (2pt);
    \fill[black] (B1) circle (2pt);
\end{tikzpicture}}
+\Dlog \frac{R_{\begin{tikzpicture}[scale=0.6]
    \coordinate (A1) at (0,0);
    \coordinate (B1) at (1,0);
            \draw[thick, red, fill=gray, opacity=0.5] (1/2,0) ellipse (0.75cm and 0.3cm);
    \draw[thick] (A1) -- (B1);
        \draw[thick, red] (1/2,0) ellipse (0.75cm and 0.3cm);
    \fill[black] (A1) circle (2pt);
    \fill[black] (B1) circle (2pt);
\end{tikzpicture}}}{R_{\input{Figures/graph_path_2_region_12_2.tex}}}
\wedge \widetilde{\mu}_{\begin{tikzpicture}[scale=0.6]
    \coordinate (A1) at (0,0);
    \coordinate (B1) at (1,0);
    \draw[thick] (A1) -- (B1);
        \draw[thick, red] (1/2,0) ellipse (0.75cm and 0.3cm);
    \fill[black] (A1) circle (2pt);
    \fill[black] (B1) circle (2pt);
\end{tikzpicture}}\right)\nonumber\\
&+\alpha_2 \,\Dlog R_{\begin{tikzpicture}[scale=0.6]
    \coordinate (A1) at (0,0);
    \coordinate (B1) at (1,0);
    \draw[thick] (A1) -- (B1);
        \draw[thick, red] (1/2,0) ellipse (0.75cm and 0.3cm);
    \draw[thick, black, fill=gray,opacity=0.5] (1,0) ellipse (0.15cm and 0.15cm);
    \fill[black] (A1) circle (2pt);
    \fill[black] (B1) circle (2pt);
\end{tikzpicture}}
\wedge \widetilde{\mu}_{\begin{tikzpicture}[scale=0.6]
    \coordinate (A1) at (0,0);
    \coordinate (B1) at (1,0);
    \draw[thick] (A1) -- (B1);
        \draw[thick, red] (1/2,0) ellipse (0.75cm and 0.3cm);
            \draw[thick, black] (1,0) ellipse (0.15cm and 0.15cm);
    \fill[black] (A1) circle (2pt);
    \fill[black] (B1) circle (2pt);
\end{tikzpicture}}\,,\\
\Dlog u_{\begin{tikzpicture}[scale=0.6]
    \coordinate (A) at (0,0);
    \coordinate (B) at (1/2,0);
    \draw[thick] (A) -- (B);
    \fill[black] (A) circle (2pt);
    \fill[black] (B) circle (2pt);
\end{tikzpicture}} \wedge \widetilde{\mu}_{\begin{tikzpicture}[scale=0.6]
    \coordinate (A1) at (0,0);
    \coordinate (B1) at (1,0);
    \draw[thick] (A1) -- (B1);
            \draw[thick, red] (0,0) ellipse (0.15cm and 0.15cm);
            \draw[thick, red] (1,0) ellipse (0.15cm and 0.15cm);
    \fill[black] (A1) circle (2pt);
    \fill[black] (B1) circle (2pt);
\end{tikzpicture}}&= 
\left(\alpha_1\,\Dlog R_{\begin{tikzpicture}[scale=0.6]
    \coordinate (A1) at (0,0);
    \coordinate (B1) at (1,0);
    \draw[thick] (A1) -- (B1);
            \draw[thick, red,fill=gray,opacity=0.5] (0,0) ellipse (0.15cm and 0.15cm);
            \draw[thick, red] (1,0) ellipse (0.15cm and 0.15cm);
    \fill[black] (A1) circle (2pt);
    \fill[black] (B1) circle (2pt);
\end{tikzpicture}}+\alpha_2\,\Dlog R_{\begin{tikzpicture}[scale=0.6]
    \coordinate (A1) at (0,0);
    \coordinate (B1) at (1,0);
    \draw[thick] (A1) -- (B1);
            \draw[thick, red,fill=gray,opacity=0.5] (1,0) ellipse (0.15cm and 0.15cm);
            \draw[thick, red] (0,0) ellipse (0.15cm and 0.15cm);
    \fill[black] (A1) circle (2pt);
    \fill[black] (B1) circle (2pt);
\end{tikzpicture}}\right)
\wedge \widetilde{\mu}_{\begin{tikzpicture}[scale=0.6]
    \coordinate (A1) at (0,0);
    \coordinate (B1) at (1,0);
    \draw[thick] (A1) -- (B1);
            \draw[thick, red] (0,0) ellipse (0.15cm and 0.15cm);
            \draw[thick, red] (1,0) ellipse (0.15cm and 0.15cm);
    \fill[black] (A1) circle (2pt);
    \fill[black] (B1) circle (2pt);
\end{tikzpicture}}\,,
\end{align}
in agreement with formula \eqref{eq:actionDlog}.
Then the matrices associated to the action of $\Dlog u_{\begin{tikzpicture}[scale=0.6]
    \coordinate (A) at (0,0);
    \coordinate (B) at (1/2,0);
    \draw[thick] (A) -- (B);
    \fill[black] (A) circle (2pt);
    \fill[black] (B) circle (2pt);
\end{tikzpicture}}$
\begin{equation}
\Dlog u_{\begin{tikzpicture}[scale=0.6]
    \coordinate (A) at (0,0);
    \coordinate (B) at (1/2,0);
    \draw[thick] (A) -- (B);
    \fill[black] (A) circle (2pt);
    \fill[black] (B) circle (2pt);
\end{tikzpicture}}\wedge v_{\begin{tikzpicture}[scale=0.6]
    \coordinate (A) at (0,0);
    \coordinate (B) at (1/2,0);
    \draw[thick] (A) -- (B);
    \fill[black] (A) circle (2pt);
    \fill[black] (B) circle (2pt);
\end{tikzpicture}}=B_{\begin{tikzpicture}[scale=0.6]
    \coordinate (A) at (0,0);
    \coordinate (B) at (1/2,0);
    \draw[thick] (A) -- (B);
    \fill[black] (A) circle (2pt);
    \fill[black] (B) circle (2pt);
\end{tikzpicture}}\wedge v_{\begin{tikzpicture}[scale=0.6]
    \coordinate (A) at (0,0);
    \coordinate (B) at (1/2,0);
    \draw[thick] (A) -- (B);
    \fill[black] (A) circle (2pt);
    \fill[black] (B) circle (2pt);
\end{tikzpicture}}=\alpha_1 B_{\begin{tikzpicture}[scale=0.6]
    \coordinate (A) at (0,0);
    \coordinate (B) at (1/2,0);
    \draw[thick] (A) -- (B);
    \fill[black] (A) circle (2pt);
    \fill[black] (B) circle (2pt);
\end{tikzpicture}}^{(1)}\wedge v_{\begin{tikzpicture}[scale=0.6]
    \coordinate (A) at (0,0);
    \coordinate (B) at (1/2,0);
    \draw[thick] (A) -- (B);
    \fill[black] (A) circle (2pt);
    \fill[black] (B) circle (2pt);
\end{tikzpicture}}+\alpha_2 B_{\begin{tikzpicture}[scale=0.6]
    \coordinate (A) at (0,0);
    \coordinate (B) at (1/2,0);
    \draw[thick] (A) -- (B);
    \fill[black] (A) circle (2pt);
    \fill[black] (B) circle (2pt);
\end{tikzpicture}}^{(2)}\wedge v_{\begin{tikzpicture}[scale=0.6]
    \coordinate (A) at (0,0);
    \coordinate (B) at (1/2,0);
    \draw[thick] (A) -- (B);
    \fill[black] (A) circle (2pt);
    \fill[black] (B) circle (2pt);
\end{tikzpicture}}\,,
\end{equation}
can be found to be
\begin{equation}
B_{\begin{tikzpicture}[scale=0.6]
    \coordinate (A) at (0,0);
    \coordinate (B) at (1/2,0);
    \draw[thick] (A) -- (B);
    \fill[black] (A) circle (2pt);
    \fill[black] (B) circle (2pt);
\end{tikzpicture}}^{(1)}=\begin{pmatrix}
\Dlog (X_1+X_2)&0&0&0\\
0&\Dlog (X_1+Y_{1,2})&0&0\\
\Dlog\frac{X_1+X_2}{X_1-Y_{1,2}}&0&\Dlog (X_1-Y_{1,2})&0\\
0&0&0&\Dlog (X_1+Y_{1,2})
\end{pmatrix}
\end{equation}
and
\begin{equation}
B_{\begin{tikzpicture}[scale=0.6]
    \coordinate (A) at (0,0);
    \coordinate (B) at (1/2,0);
    \draw[thick] (A) -- (B);
    \fill[black] (A) circle (2pt);
    \fill[black] (B) circle (2pt);
\end{tikzpicture}}^{(2)}=\begin{pmatrix}
\Dlog (X_1+X_2)&0&0&0\\
\Dlog\frac{X_1+X_2}{X_2-Y_{1,2}}&\Dlog (X_2-Y_{1,2})&0&0\\
0&0&\Dlog (X_2+Y_{1,2})&0\\
0&0&0&\Dlog (X_2+Y_{1,2})
\end{pmatrix}
\end{equation}
It is easy to check that the matrices $B_{\begin{tikzpicture}[scale=0.6]
    \coordinate (A) at (0,0);
    \coordinate (B) at (1/2,0);
    \draw[thick] (A) -- (B);
    \fill[black] (A) circle (2pt);
    \fill[black] (B) circle (2pt);
\end{tikzpicture}}^{(1)}$ and $B_{\begin{tikzpicture}[scale=0.6]
    \coordinate (A) at (0,0);
    \coordinate (B) at (1/2,0);
    \draw[thick] (A) -- (B);
    \fill[black] (A) circle (2pt);
    \fill[black] (B) circle (2pt);
\end{tikzpicture}}^{(2)}$ anticommute. The diagonalisation of matrix $B_{\begin{tikzpicture}[scale=0.6]
    \coordinate (A) at (0,0);
    \coordinate (B) at (1/2,0);
    \draw[thick] (A) -- (B);
    \fill[black] (A) circle (2pt);
    \fill[black] (B) circle (2pt);
\end{tikzpicture}}$ provides a new set of forms, labelled by the same $c$-tubings
\begin{equation}
\bar{v}_{\begin{tikzpicture}[scale=0.6]
    \coordinate (A) at (0,0);
    \coordinate (B) at (1/2,0);
    \draw[thick] (A) -- (B);
    \fill[black] (A) circle (2pt);
    \fill[black] (B) circle (2pt);
\end{tikzpicture}}=\left\{\overline{\mu}_{\begin{tikzpicture}[scale=0.6]
    \coordinate (A1) at (0,0);
    \coordinate (B1) at (1,0);
    \draw[thick] (A1) -- (B1);
        \draw[thick, red] (1/2,0) ellipse (0.75cm and 0.3cm);
    \fill[black] (A1) circle (2pt);
    \fill[black] (B1) circle (2pt);
\end{tikzpicture}},\overline{\mu}_{\begin{tikzpicture}[scale=0.6]
    \coordinate (A1) at (0,0);
    \coordinate (B1) at (1,0);
    \draw[thick] (A1) -- (B1);
        \draw[thick, red] (1/2,0) ellipse (0.75cm and 0.3cm);
            \draw[thick, black] (0,0) ellipse (0.15cm and 0.15cm);
    \fill[black] (A1) circle (2pt);
    \fill[black] (B1) circle (2pt);
\end{tikzpicture}},\overline{\mu}_{\begin{tikzpicture}[scale=0.6]
    \coordinate (A1) at (0,0);
    \coordinate (B1) at (1,0);
    \draw[thick] (A1) -- (B1);
        \draw[thick, red] (1/2,0) ellipse (0.75cm and 0.3cm);
            \draw[thick, black] (1,0) ellipse (0.15cm and 0.15cm);
    \fill[black] (A1) circle (2pt);
    \fill[black] (B1) circle (2pt);
\end{tikzpicture}},\overline{\mu}_{\begin{tikzpicture}[scale=0.6]
    \coordinate (A1) at (0,0);
    \coordinate (B1) at (1,0);
    \draw[thick] (A1) -- (B1);
            \draw[thick, red] (0,0) ellipse (0.15cm and 0.15cm);
            \draw[thick, red] (1,0) ellipse (0.15cm and 0.15cm);
    \fill[black] (A1) circle (2pt);
    \fill[black] (B1) circle (2pt);
\end{tikzpicture}}\right\},
\end{equation}
where
\begin{align}
\overline{\mu}_{\begin{tikzpicture}[scale=0.6]
    \coordinate (A1) at (0,0);
    \coordinate (B1) at (1,0);
    \draw[thick] (A1) -- (B1);
        \draw[thick, red] (1/2,0) ellipse (0.75cm and 0.3cm);
    \fill[black] (A1) circle (2pt);
    \fill[black] (B1) circle (2pt);
\end{tikzpicture}}&=\widetilde{\mu}_{\begin{tikzpicture}[scale=0.6]
    \coordinate (A1) at (0,0);
    \coordinate (B1) at (1,0);
    \draw[thick] (A1) -- (B1);
        \draw[thick, red] (1/2,0) ellipse (0.75cm and 0.3cm);
    \fill[black] (A1) circle (2pt);
    \fill[black] (B1) circle (2pt);
\end{tikzpicture}}\,,\\
\overline{\mu}_{\begin{tikzpicture}[scale=0.6]
    \coordinate (A1) at (0,0);
    \coordinate (B1) at (1,0);
    \draw[thick] (A1) -- (B1);
        \draw[thick, red] (1/2,0) ellipse (0.75cm and 0.3cm);
            \draw[thick, black] (0,0) ellipse (0.15cm and 0.15cm);
    \fill[black] (A1) circle (2pt);
    \fill[black] (B1) circle (2pt);
\end{tikzpicture}}&=\left(\frac{\alpha_1( X_2-Y_{1,2})}{\alpha_2(X_1+Y_{1,2})}-1\right)\widetilde{\mu}_{\begin{tikzpicture}[scale=0.6]
    \coordinate (A1) at (0,0);
    \coordinate (B1) at (1,0);
    \draw[thick] (A1) -- (B1);
        \draw[thick, red] (1/2,0) ellipse (0.75cm and 0.3cm);
            \draw[thick, black] (0,0) ellipse (0.15cm and 0.15cm);
    \fill[black] (A1) circle (2pt);
    \fill[black] (B1) circle (2pt);
\end{tikzpicture}}+\widetilde{\mu}_{\begin{tikzpicture}[scale=0.6]
    \coordinate (A1) at (0,0);
    \coordinate (B1) at (1,0);
    \draw[thick] (A1) -- (B1);
        \draw[thick, red] (1/2,0) ellipse (0.75cm and 0.3cm);
    \fill[black] (A1) circle (2pt);
    \fill[black] (B1) circle (2pt);
\end{tikzpicture}}\,,\\
\overline{\mu}_{\begin{tikzpicture}[scale=0.6]
    \coordinate (A1) at (0,0);
    \coordinate (B1) at (1,0);
    \draw[thick] (A1) -- (B1);
        \draw[thick, red] (1/2,0) ellipse (0.75cm and 0.3cm);
            \draw[thick, black] (1,0) ellipse (0.15cm and 0.15cm);
    \fill[black] (A1) circle (2pt);
    \fill[black] (B1) circle (2pt);
\end{tikzpicture}}&=\left(\frac{\alpha_2(X_1-Y_{1,2})}{\alpha_1(X_2+Y_{1,2})}-1\right)\widetilde{\mu}_{\begin{tikzpicture}[scale=0.6]
    \coordinate (A1) at (0,0);
    \coordinate (B1) at (1,0);
    \draw[thick] (A1) -- (B1);
        \draw[thick, red] (1/2,0) ellipse (0.75cm and 0.3cm);
            \draw[thick, black] (1,0) ellipse (0.15cm and 0.15cm);
    \fill[black] (A1) circle (2pt);
    \fill[black] (B1) circle (2pt);
\end{tikzpicture}}+\widetilde{\mu}_{\begin{tikzpicture}[scale=0.6]
    \coordinate (A1) at (0,0);
    \coordinate (B1) at (1,0);
    \draw[thick] (A1) -- (B1);
        \draw[thick, red] (1/2,0) ellipse (0.75cm and 0.3cm);
    \fill[black] (A1) circle (2pt);
    \fill[black] (B1) circle (2pt);
\end{tikzpicture}}\,,\\
\overline{\mu}_{\begin{tikzpicture}[scale=0.6]
    \coordinate (A1) at (0,0);
    \coordinate (B1) at (1,0);
    \draw[thick] (A1) -- (B1);
            \draw[thick, red] (0,0) ellipse (0.15cm and 0.15cm);
            \draw[thick, red] (1,0) ellipse (0.15cm and 0.15cm);
    \fill[black] (A1) circle (2pt);
    \fill[black] (B1) circle (2pt);
\end{tikzpicture}}&=\widetilde{\mu}_{\begin{tikzpicture}[scale=0.6]
    \coordinate (A1) at (0,0);
    \coordinate (B1) at (1,0);
    \draw[thick] (A1) -- (B1);
            \draw[thick, red] (0,0) ellipse (0.15cm and 0.15cm);
            \draw[thick, red] (1,0) ellipse (0.15cm and 0.15cm);
    \fill[black] (A1) circle (2pt);
    \fill[black] (B1) circle (2pt);
\end{tikzpicture}}\,,
\end{align}
in agreement with formula \eqref{eq:eigenvector}.
When acting with $\Dlog u_{\begin{tikzpicture}[scale=0.6]
    \coordinate (A) at (0,0);
    \coordinate (B) at (1/2,0);
    \draw[thick] (A) -- (B);
    \fill[black] (A) circle (2pt);
    \fill[black] (B) circle (2pt);
\end{tikzpicture}}$ on the elements of $\bar{v}_{\begin{tikzpicture}[scale=0.6]
    \coordinate (A) at (0,0);
    \coordinate (B) at (1/2,0);
    \draw[thick] (A) -- (B);
    \fill[black] (A) circle (2pt);
    \fill[black] (B) circle (2pt);
\end{tikzpicture}}$, we calculate the eigenvalues of matrix $B_{\begin{tikzpicture}[scale=0.6]
    \coordinate (A) at (0,0);
    \coordinate (B) at (1/2,0);
    \draw[thick] (A) -- (B);
    \fill[black] (A) circle (2pt);
    \fill[black] (B) circle (2pt);
\end{tikzpicture}}$ to be
\begin{align}
\Dlog u_{\begin{tikzpicture}[scale=0.6]
    \coordinate (A) at (0,0);
    \coordinate (B) at (1/2,0);
    \draw[thick] (A) -- (B);
    \fill[black] (A) circle (2pt);
    \fill[black] (B) circle (2pt);
\end{tikzpicture}} \wedge \overline{\mu}_{\begin{tikzpicture}[scale=0.6]
    \coordinate (A1) at (0,0);
    \coordinate (B1) at (1,0);
    \draw[thick] (A1) -- (B1);
        \draw[thick, red] (1/2,0) ellipse (0.75cm and 0.3cm);
    \fill[black] (A1) circle (2pt);
    \fill[black] (B1) circle (2pt);
\end{tikzpicture}}&= 
(\alpha_1+\alpha_2)\Dlog R_{\begin{tikzpicture}[scale=0.6]
    \coordinate (A1) at (0,0);
    \coordinate (B1) at (1,0);
            \draw[thick, red, fill=gray, opacity=0.5] (1/2,0) ellipse (0.75cm and 0.3cm);
    \draw[thick] (A1) -- (B1);
        \draw[thick, red] (1/2,0) ellipse (0.75cm and 0.3cm);
    \fill[black] (A1) circle (2pt);
    \fill[black] (B1) circle (2pt);
\end{tikzpicture}}
\wedge \overline{\mu}_{\begin{tikzpicture}[scale=0.6]
    \coordinate (A1) at (0,0);
    \coordinate (B1) at (1,0);
    \draw[thick] (A1) -- (B1);
        \draw[thick, red] (1/2,0) ellipse (0.75cm and 0.3cm);
    \fill[black] (A1) circle (2pt);
    \fill[black] (B1) circle (2pt);
\end{tikzpicture}}\,,\\
\Dlog u_{\begin{tikzpicture}[scale=0.6]
    \coordinate (A) at (0,0);
    \coordinate (B) at (1/2,0);
    \draw[thick] (A) -- (B);
    \fill[black] (A) circle (2pt);
    \fill[black] (B) circle (2pt);
\end{tikzpicture}} \wedge \overline{\mu}_{\begin{tikzpicture}[scale=0.6]
    \coordinate (A1) at (0,0);
    \coordinate (B1) at (1,0);
    \draw[thick] (A1) -- (B1);
        \draw[thick, red] (1/2,0) ellipse (0.75cm and 0.3cm);
            \draw[thick, black] (0,0) ellipse (0.15cm and 0.15cm);
    \fill[black] (A1) circle (2pt);
    \fill[black] (B1) circle (2pt);
\end{tikzpicture}}&= 
(\alpha_1\Dlog R_{\begin{tikzpicture}[scale=0.6]
    \coordinate (A1) at (0,0);
    \coordinate (B1) at (1,0);
    \draw[thick] (A1) -- (B1);

    \draw[thick, black, fill=gray,opacity=0.5] (0,0) ellipse (0.15cm and 0.15cm);
            \draw[thick, red] (1/2,0) ellipse (0.75cm and 0.3cm);
    \fill[black] (A1) circle (2pt);
    \fill[black] (B1) circle (2pt);
\end{tikzpicture}}+\alpha_2\Dlog R_{\input{Figures/graph_path_2_region_12_1.tex}})
\wedge \overline{\mu}_{\begin{tikzpicture}[scale=0.6]
    \coordinate (A1) at (0,0);
    \coordinate (B1) at (1,0);
    \draw[thick] (A1) -- (B1);
        \draw[thick, red] (1/2,0) ellipse (0.75cm and 0.3cm);
            \draw[thick, black] (0,0) ellipse (0.15cm and 0.15cm);
    \fill[black] (A1) circle (2pt);
    \fill[black] (B1) circle (2pt);
\end{tikzpicture}}\,,\\
\Dlog u_{\begin{tikzpicture}[scale=0.6]
    \coordinate (A) at (0,0);
    \coordinate (B) at (1/2,0);
    \draw[thick] (A) -- (B);
    \fill[black] (A) circle (2pt);
    \fill[black] (B) circle (2pt);
\end{tikzpicture}} \wedge \overline{\mu}_{\begin{tikzpicture}[scale=0.6]
    \coordinate (A1) at (0,0);
    \coordinate (B1) at (1,0);
    \draw[thick] (A1) -- (B1);
        \draw[thick, red] (1/2,0) ellipse (0.75cm and 0.3cm);
            \draw[thick, black] (1,0) ellipse (0.15cm and 0.15cm);
    \fill[black] (A1) circle (2pt);
    \fill[black] (B1) circle (2pt);
\end{tikzpicture}}&= 
(\alpha_1\Dlog R_{\input{Figures/graph_path_2_region_12_2.tex}}+\alpha_2\Dlog R_{\begin{tikzpicture}[scale=0.6]
    \coordinate (A1) at (0,0);
    \coordinate (B1) at (1,0);
    \draw[thick] (A1) -- (B1);
        \draw[thick, red] (1/2,0) ellipse (0.75cm and 0.3cm);
    \draw[thick, black, fill=gray,opacity=0.5] (1,0) ellipse (0.15cm and 0.15cm);
    \fill[black] (A1) circle (2pt);
    \fill[black] (B1) circle (2pt);
\end{tikzpicture}})
\wedge \overline{\mu}_{\begin{tikzpicture}[scale=0.6]
    \coordinate (A1) at (0,0);
    \coordinate (B1) at (1,0);
    \draw[thick] (A1) -- (B1);
        \draw[thick, red] (1/2,0) ellipse (0.75cm and 0.3cm);
            \draw[thick, black] (1,0) ellipse (0.15cm and 0.15cm);
    \fill[black] (A1) circle (2pt);
    \fill[black] (B1) circle (2pt);
\end{tikzpicture}}\,,\\
\Dlog u_{\begin{tikzpicture}[scale=0.6]
    \coordinate (A) at (0,0);
    \coordinate (B) at (1/2,0);
    \draw[thick] (A) -- (B);
    \fill[black] (A) circle (2pt);
    \fill[black] (B) circle (2pt);
\end{tikzpicture}} \wedge \overline{\mu}_{\begin{tikzpicture}[scale=0.6]
    \coordinate (A1) at (0,0);
    \coordinate (B1) at (1,0);
    \draw[thick] (A1) -- (B1);
            \draw[thick, red] (0,0) ellipse (0.15cm and 0.15cm);
            \draw[thick, red] (1,0) ellipse (0.15cm and 0.15cm);
    \fill[black] (A1) circle (2pt);
    \fill[black] (B1) circle (2pt);
\end{tikzpicture}}&= 
(\alpha_1\Dlog R_{\begin{tikzpicture}[scale=0.6]
    \coordinate (A1) at (0,0);
    \coordinate (B1) at (1,0);
    \draw[thick] (A1) -- (B1);
            \draw[thick, red,fill=gray,opacity=0.5] (0,0) ellipse (0.15cm and 0.15cm);
            \draw[thick, red] (1,0) ellipse (0.15cm and 0.15cm);
    \fill[black] (A1) circle (2pt);
    \fill[black] (B1) circle (2pt);
\end{tikzpicture}}+\alpha_2\Dlog R_{\begin{tikzpicture}[scale=0.6]
    \coordinate (A1) at (0,0);
    \coordinate (B1) at (1,0);
    \draw[thick] (A1) -- (B1);
            \draw[thick, red,fill=gray,opacity=0.5] (1,0) ellipse (0.15cm and 0.15cm);
            \draw[thick, red] (0,0) ellipse (0.15cm and 0.15cm);
    \fill[black] (A1) circle (2pt);
    \fill[black] (B1) circle (2pt);
\end{tikzpicture}})
\wedge \overline{\mu}_{\begin{tikzpicture}[scale=0.6]
    \coordinate (A1) at (0,0);
    \coordinate (B1) at (1,0);
    \draw[thick] (A1) -- (B1);
            \draw[thick, red] (0,0) ellipse (0.15cm and 0.15cm);
            \draw[thick, red] (1,0) ellipse (0.15cm and 0.15cm);
    \fill[black] (A1) circle (2pt);
    \fill[black] (B1) circle (2pt);
\end{tikzpicture}}\,.
\end{align}
The terms multiplying $\alpha_v$ are just the $\Dlog$ of the region variables of the  region containing vertex $v$, in agreement with formula \eqref{eq:eigenvalue}.

\newpage
\vspace{0.3cm}
\noindent{$\bullet$ \bf Path graph \begin{tikzpicture}[scale=0.6]
    \coordinate (A) at (0,0);
    \coordinate (B) at (1/2,0);
        \coordinate (C) at (1,0);
    \draw[thick] (A) -- (B) -- (C);
    \fill[black] (A) circle (2pt);
    \fill[black] (B) circle (2pt);
        \fill[black] (C) circle (2pt);
\end{tikzpicture}}

The next tree graph that we consider is the path graph with three vertices:
\begin{equation}
\begin{tikzpicture}[scale=1.2]
    \coordinate (A) at (0,0);
    \coordinate (B) at (1,0);
        \coordinate (C) at (2,0);
                \coordinate (E) at (-1,0);
    \draw[thick] (A) -- (B) -- (C);
    \fill[black] (A) circle (2pt) node[above] {$X_1$};
    \fill[black] (B) circle (2pt) node[above] {$X_2$};
        \fill[black] (C) circle (2pt) node[above] {$X_3$};
    \node at ($(A)!0.5!(B)$) [below] {$Y_{1,2}$};
        \node at ($(B)!0.5!(C)$) [below] {$Y_{2,3}$};
            \node at (E) {$G=$};
\end{tikzpicture}
\end{equation}
The main purpose of this example is to highlight the differences between our method and the one based on the kinematic flow \cite{Baumann:2025qjx}. The two methods generically differ in the number and the labelling of the master functions, and the path graph with three vertices is the first example that will illustrate it. We will not provide an analysis of this case as detailed as we did for the path graph with two vertices.

The set of differential forms that we consider is labelled by all $c$-tubings of $G$ and the flat space canonical form $\widetilde{\Omega}^{\text{flat}}_G$ is decomposed as follows
\begin{align}
\widetilde{\Omega}^{\text{flat}}_{\begin{tikzpicture}[scale=0.6]
    \coordinate (A) at (0,0);
    \coordinate (B) at (1/2,0);
        \coordinate (C) at (1,0);
    \draw[thick] (A) -- (B) -- (C);
    \fill[black] (A) circle (2pt);
    \fill[black] (B) circle (2pt);
        \fill[black] (C) circle (2pt);
\end{tikzpicture}}(X_v+x_v,Y_e)&=\Big(
\widetilde{\mu}_{\input{Figures/graph_path_3_u_tubing_1_12_123.tex}}+\widetilde{\mu}_{\input{Figures/graph_path_3_u_tubing_12_2_123.tex}}+
\widetilde{\mu}_{\input{Figures/graph_path_3_u_tubing_123_2_23.tex}}
+\widetilde{\mu}_{\input{Figures/graph_path_3_u_tubing_123_23_3.tex}}+\widetilde{\mu}_{\input{Figures/graph_path_3_u_tubing_1_123_3.tex}}\nonumber\\
&-\widetilde{\mu}_{\input{Figures/graph_path_3_u_tubing_1_123_123.tex}}-\widetilde{\mu}_{\input{Figures/graph_path_3_u_tubing_123_2_123.tex}}-
\widetilde{\mu}_{\input{Figures/graph_path_3_u_tubing_123_123_3.tex}}
-\widetilde{\mu}_{\input{Figures/graph_path_3_u_tubing_12_12_123.tex}}-\widetilde{\mu}_{\input{Figures/graph_path_3_u_tubing_123_23_23.tex}}+\widetilde{\mu}_{\begin{tikzpicture}[scale=0.4]
    \coordinate (A1) at (0,0);
    \coordinate (B1) at (1,0);
        \coordinate (C1) at (2,0);
    \draw[thick] (A1) -- (B1)-- (C1);
            \draw[thick, red] (1,0) ellipse (1.5cm and 0.6cm);

    \fill[black] (A1) circle (2pt);
    \fill[black] (B1) circle (2pt);
        \fill[black] (C1) circle (2pt);
\end{tikzpicture}} \Big)\nonumber\\
&-\Big( \widetilde{\mu}_{\input{Figures/graph_path_3_u_tubing_1_12_3.tex}}+\widetilde{\mu}_{\input{Figures/graph_path_3_u_tubing_12_2_3.tex}}-\widetilde{\mu}_{\input{Figures/graph_path_3_u_tubing_12_12_3.tex}}\Big)-\Big( \widetilde{\mu}_{\input{Figures/graph_path_3_u_tubing_1_2_23.tex}}+\widetilde{\mu}_{\input{Figures/graph_path_3_u_tubing_1_23_3.tex}}-\widetilde{\mu}_{\input{Figures/graph_path_3_u_tubing_1_23_23.tex}}\Big)\nonumber\\
&+\widetilde{\mu}_{\input{Figures/graph_path_3_u_tubing_1_2_3.tex}}\,.
\end{align}
The total number of functions we consider for this graph is $18$, which differs from the $16$ basis elements in the kinematic flow approach. The increase in the number of functions is the price for keeping all differential forms logarithmic, a price that we are willing to pay. 

The action of the $\Dlog u$ on the 18-dimensional space splits into blocks of size 11,3,3 and 1, where we have already encountered the latter three blocks in our study of the path graph with two vertices. Below we provide the action of $\Dlog u$ on one of the 11 function in the largest block:
\begin{align}
\Dlog u \wedge \widetilde{\mu}_{\input{Figures/graph_path_3_u_tubing_1_12_123.tex}}&= \alpha_1 \Dlog(X_1+Y_{1,2})\wedge \widetilde{\mu}_{\input{Figures/graph_path_3_u_tubing_1_12_123.tex}}\nonumber\\
&+\alpha_2\left(\Dlog(X_2-Y_{1,2}+Y_{2,3})\wedge \widetilde{\mu}_{\input{Figures/graph_path_3_u_tubing_1_12_123.tex}}+\Dlog\frac{X_1+X_2+Y_{2,3}}{X_2-Y_{1,2}+Y_{2,3}}\wedge \widetilde{\mu}_{\input{Figures/graph_path_3_u_tubing_12_12_123.tex}}\right)\nonumber\\
&+\alpha_3\Big(\Dlog(X_3-Y_{2,3})\wedge \widetilde{\mu}_{\input{Figures/graph_path_3_u_tubing_1_12_123.tex}}+\Dlog\frac{X_2+X_3-Y_{1,2}}{X_3-Y_{2,3}}\wedge \widetilde{\mu}_{\input{Figures/graph_path_3_u_tubing_1_123_123.tex}}\nonumber\\
&+\Dlog\frac{X_1+X_2+X_{3}}{X_2+X_3-Y_{1,2}}\wedge \widetilde{\mu}_{\begin{tikzpicture}[scale=0.4]
    \coordinate (A1) at (0,0);
    \coordinate (B1) at (1,0);
        \coordinate (C1) at (2,0);
    \draw[thick] (A1) -- (B1)-- (C1);
            \draw[thick, red] (1,0) ellipse (1.5cm and 0.6cm);

    \fill[black] (A1) circle (2pt);
    \fill[black] (B1) circle (2pt);
        \fill[black] (C1) circle (2pt);
\end{tikzpicture}}\Big) \,,
\end{align}
where all arguments in the $\Dlog$'s can be identified with appropriate region variables. We also provide an explicit form of the eigenvector of the matrix of coefficients in our differential equation associated to this $c$-tubing
\begin{align}
\overline{\mu}_{\input{Figures/graph_path_3_u_tubing_1_12_123.tex}}&=\left(\frac{\alpha_1(X_2-Y_{1,2}+Y_{2,3})}{\alpha_2(X_1+Y_{1,2})}-1\right)\left(\frac{\alpha_2(X_3-Y_{2,3})}{\alpha_3(X_2-Y_{1,2}+Y_{2,3})}-1\right)\overline{\mu}_{\input{Figures/graph_path_3_u_tubing_1_12_123.tex}}+\overline{\mu}_{\begin{tikzpicture}[scale=0.4]
    \coordinate (A1) at (0,0);
    \coordinate (B1) at (1,0);
        \coordinate (C1) at (2,0);
    \draw[thick] (A1) -- (B1)-- (C1);
            \draw[thick, red] (1,0) ellipse (1.5cm and 0.6cm);

    \fill[black] (A1) circle (2pt);
    \fill[black] (B1) circle (2pt);
        \fill[black] (C1) circle (2pt);
\end{tikzpicture}}\nonumber\\
&+\left(\frac{\alpha_1(X_2-Y_{1,2}+Y_{2,3})}{\alpha_2(X_1+Y_{1,2})}-1\right)\overline{\mu}_{\input{Figures/graph_path_3_u_tubing_1_123_123.tex}}+\left(\frac{\alpha_2(X_3-Y_{2,3})}{\alpha_3(X_2-Y_{1,2}+Y_{2,3})}-1\right)\overline{\mu}_{\input{Figures/graph_path_3_u_tubing_12_12_123.tex}}\,,
\end{align}
and an explicit form of the action of $\Dlog$ on the eigenvector
\begin{align}
\Dlog u \wedge\overline{\mu}_{\input{Figures/graph_path_3_u_tubing_1_12_123.tex}}&=(\alpha_1 \Dlog(X_1+Y_{1,2})+\alpha_2 \Dlog (X_2-Y_{1,2}+Y_{2,3})\nonumber\\
&+\alpha_3 \Dlog(X_3-Y_{2,3}))\wedge\overline{\mu}_{\input{Figures/graph_path_3_u_tubing_1_12_123.tex}}\\
&=(\alpha_1 \Dlog R_{\input{Figures/graph_path_3_u_region_1.tex}}+\alpha_2 \Dlog R_{\input{Figures/graph_path_3_u_region_1_12.tex}}
+\alpha_3 \Dlog R_{\input{Figures/graph_path_3_u_region_12_123.tex}})\wedge\overline{\mu}_{\input{Figures/graph_path_3_u_tubing_1_12_123.tex}}\,.
\end{align}
All these results agree with the general formulae presented in section \ref{sec:diffeqs}.
 
 \vspace{0.3cm}
\noindent{$\bullet$ \bf Star graph \raisebox{-0.2cm}{\input{Figures/graph_star_4p.tex}}}

Finally, we want to illustrate an explicit form of the differential equations for a more complicated example. Let us consider the star graph
\begin{equation}
\begin{tikzpicture}[scale=0.8]
    \coordinate (A) at (0,0);
    \coordinate (B) at (0,1);
        \coordinate (C) at (-2/3-0.2,-2/3);
                \coordinate (D) at (2/3+0.2,-2/3);
    \draw[thick] (B) -- (A) -- (C);
        \draw[thick] (D) -- (A) ;
    \fill[black] (A) circle (2pt) node[right] {$X_1$};
        \fill[black] (B) circle (2pt) node[above] {$X_2$};
            \fill[black] (C) circle (2pt) node[below] {$X_3$};
                \fill[black] (D) circle (2pt) node[below] {$X_4$};
\end{tikzpicture}
\end{equation}
We consider the following differential form
\begin{align}
\widetilde{\mu}_{\tau_1}&=\dlog \frac{x_1+X_1+Y_{1,2}+Y_{1,3}+Y_{1,4}}{x_1}\wedge\dlog \frac{x_1+X_1+x_2+X_2+Y_{1,3}+Y_{1,4}}{x_2}\nonumber\\ 
&\wedge\dlog \frac{x_1+X_1+x_2+X_2+x_3+X_3+Y_{1,4}}{x_3}\wedge \dlog \frac{x_4+X_4+Y_{1,4}}{x_4}\,,
\end{align}
where we defined
\begin{equation}
\tau_{1}=\raisebox{-0.5cm}{\input{Figures/graph_star_4_tubing_1_12_123_4.tex}}\qquad\tau_{2}=\raisebox{-0.5cm}{\input{Figures/graph_star_4_tubing_1_123_123_4.tex}}\qquad\tau_{3}=\raisebox{-0.5cm}{\input{Figures/graph_star_4_tubing_12_12_123_4.tex}}\qquad\tau_4=\raisebox{-0.5cm}{\input{Figures/graph_star_4_tubing_123_123_123_4.tex}}
\end{equation}
The action of $\Dlog u$ with $u=x_1^{\alpha_1}x_2^{\alpha_2}x_3^{\alpha_3}x_4^{\alpha_4}$ on this function is then
\begin{align}
\Dlog u\wedge\widetilde{\mu}_{\tau_1}&=\alpha_1 \Dlog (X_1+Y_{1,2}+Y_{1,3}+Y_{1,4})\widetilde{\mu}_{\tau_1}\nonumber\\
&+\alpha_2\left( \Dlog (X_2-Y_{1,2})\wedge\widetilde{\mu}_{\tau_1}+\Dlog \frac{X_1+X_2+Y_{1,3}+Y_{1,4}}{X_2-Y_{1,2}}\wedge\widetilde{\mu}_{\tau_3}\right)\nonumber\\
&+\alpha_3\Big( \Dlog (X_3-Y_{1,3})\wedge\widetilde{\mu}_{\tau_1}+\Dlog \frac{X_2+X_3-Y_{1,2}-Y_{1,3}}{X_3-Y_{1,3}}\wedge\widetilde{\mu}_{\tau_2}\nonumber\\
&+\Dlog \frac{X_1+X_2+X_3+Y_{1,4}}{X_2+X_3-Y_{1,2}-Y_{1,3}}\wedge\widetilde{\mu}_{\tau_4}\Big)\nonumber\\
&+\alpha_4 \Dlog (X_4+Y_{1,4})\wedge\widetilde{\mu}_{\tau_1}\,.
\end{align}
The eigenvector associated to the tubing $\tau_1$ is
\begin{align}
\overline{\mu}_{\tau_1}&=\left(\frac{\alpha_1(X_2-Y_{1,2})}{\alpha_2(X_1+Y_{1,2}+Y_{1,3}+Y_{1,4})}-1\right)\left(\frac{\alpha_2(X_3-Y_{1,3})}{\alpha_3(X_2-Y_{1,2})}-1\right)\widetilde{\mu}_{\tau_1}+\widetilde{\mu}_{\tau_4}\nonumber\\
&+\left(\frac{\alpha_1(X_2-Y_{1,2})}{\alpha_2(X_1+Y_{1,2}+Y_{1,3}+Y_{1,4})}-1\right)\widetilde{\mu}_{\tau_2}+\left(\frac{\alpha_2(X_3-Y_{1,3})}{\alpha_3(X_2-Y_{1,2})}-1\right)\widetilde{\mu}_{\tau_3}\,,
\end{align}
with the action of $\Dlog$ given by
\begin{align}
\Dlog u\wedge\overline{\mu}_{\tau_1}&=\Big(\alpha_1 \Dlog(X_1+Y_{1,2}+Y_{1,3}+Y_{1,4})+\alpha_2 \Dlog(X_2-Y_{1,2})\nonumber\\&+\alpha_3\Dlog(X_3-Y_{1,3})+\alpha_4\Dlog(X_4+Y_{1,4})\Big)\wedge\overline{\mu}_{\tau_1}\,.
\end{align}

\subsection{Graphs with Cycles}
There is no significant difference when our method is applied to graphs with cycles. To illustrate it, we present here the details for the graph
\begin{equation}
G=\raisebox{-1cm}{\begin{tikzpicture}[scale=1.4]
    \coordinate (A) at (0,0);
    \coordinate (B) at (1,0);
	\draw[thick] (A) to [bend left=60] (B);
    \draw[thick] (A) to [bend right=60] (B);
    \fill[black] (A) circle (2pt);
    \fill[black] (B) circle (2pt);
    \node[font=\small] at (-0.3,0) {$X_1$};
    \node[font=\small] at (1.3,0) {$X_2$};
        \node[font=\small] at (0.5,0.5) {$Y_{1,2}^{(1)}$};
            \node[font=\small] at (0.5,-0.5) {$Y_{1,2}^{(2)}$};
\end{tikzpicture}}
\end{equation}
There are five $b$-tubes for this graph
\begin{equation}
B_{\begin{tikzpicture}[scale=0.6]
    \coordinate (A) at (0,0);
    \coordinate (B) at (1,0);
	\draw[thick] (A) to [bend left=60] (B);
    \draw[thick] (A) to [bend right=60] (B);
    \fill[black] (A) circle (2pt);
    \fill[black] (B) circle (2pt);
\end{tikzpicture}}=\{\raisebox{-0.15cm}{\begin{tikzpicture}[scale=0.6]
    \coordinate (A) at (0,0);
    \coordinate (B) at (1,0);
	\draw[thick] (A) to [bend left=60] (B);
    \draw[thick] (A) to [bend right=60] (B);
    \fill[black] (A) circle (2pt);
    \fill[black] (B) circle (2pt);
        \draw[thick, black] (0,0) ellipse (0.15cm and 0.15cm);
\end{tikzpicture}},\raisebox{-0.15cm}{\begin{tikzpicture}[scale=0.6]
    \coordinate (A) at (0,0);
    \coordinate (B) at (1,0);
	\draw[thick] (A) to [bend left=60] (B);
    \draw[thick] (A) to [bend right=60] (B);
    \fill[black] (A) circle (2pt);
    \fill[black] (B) circle (2pt);
        \draw[thick, black] (1,0) ellipse (0.15cm and 0.15cm);
\end{tikzpicture}},\raisebox{-0.20cm}{\begin{tikzpicture}[scale=0.6]
    \coordinate (A) at (0,0);
    \coordinate (B) at (1,0);
	\draw[thick] (A) to [bend left=60] (B);
    \draw[thick] (A) to [bend right=60] (B);
    \fill[black] (A) circle (2pt);
    \fill[black] (B) circle (2pt);

                \draw[thick, black] (1/2,0) ellipse (0.75cm and 0.5cm);
\end{tikzpicture}},\raisebox{-0.22cm}{\input{Figures/graph_cyclic_2_tube_12_m1.tex}},\raisebox{-0.15cm}{\input{Figures/graph_cyclic_2_tube_12_m2.tex}}\}\,,
\end{equation}
to which we can associate the tube variables $H_T$ using formula \eqref{eq:Hb}
\begin{align}
H_{\begin{tikzpicture}[scale=0.6]
    \coordinate (A) at (0,0);
    \coordinate (B) at (1,0);
	\draw[thick] (A) to [bend left=60] (B);
    \draw[thick] (A) to [bend right=60] (B);
    \fill[black] (A) circle (2pt);
    \fill[black] (B) circle (2pt);
        \draw[thick, black] (0,0) ellipse (0.15cm and 0.15cm);
\end{tikzpicture}}&=X_1+Y_{1,2}^{(1)}+Y_{1,2}^{(2)}\,,\qquad 
H_{\begin{tikzpicture}[scale=0.6]
    \coordinate (A) at (0,0);
    \coordinate (B) at (1,0);
	\draw[thick] (A) to [bend left=60] (B);
    \draw[thick] (A) to [bend right=60] (B);
    \fill[black] (A) circle (2pt);
    \fill[black] (B) circle (2pt);
        \draw[thick, black] (1,0) ellipse (0.15cm and 0.15cm);
\end{tikzpicture}}=X_2+Y_{1,2}^{(1)}+Y_{1,2}^{(2)}\,,\qquad
H_{\begin{tikzpicture}[scale=0.6]
    \coordinate (A) at (0,0);
    \coordinate (B) at (1,0);
	\draw[thick] (A) to [bend left=60] (B);
    \draw[thick] (A) to [bend right=60] (B);
    \fill[black] (A) circle (2pt);
    \fill[black] (B) circle (2pt);

                \draw[thick, black] (1/2,0) ellipse (0.75cm and 0.5cm);
\end{tikzpicture}}=X_1+X_2\,,\nonumber\\
H_{\input{Figures/graph_cyclic_2_tube_12_m1.tex}}&=X_1+X_2+2Y_{1,2}^{(1)}\,,\qquad
H_{\input{Figures/graph_cyclic_2_tube_12_m2.tex}}=X_1+X_2+2Y_{1,2}^{(2)}\,.
\end{align}
The set of functions labelled by all possible $c$-tubings is
\begin{equation}
v_{\begin{tikzpicture}[scale=0.6]
    \coordinate (A) at (0,0);
    \coordinate (B) at (1,0);
	\draw[thick] (A) to [bend left=60] (B);
    \draw[thick] (A) to [bend right=60] (B);
    \fill[black] (A) circle (2pt);
    \fill[black] (B) circle (2pt);
\end{tikzpicture}}=\left\{\widetilde{\mu}_{\begin{tikzpicture}[scale=0.6]
    \coordinate (A) at (0,0);
    \coordinate (B) at (1,0);
	\draw[thick] (A) to [bend left=60] (B);
    \draw[thick] (A) to [bend right=60] (B);
    \fill[black] (A) circle (2pt);
    \fill[black] (B) circle (2pt);

                \draw[thick, red] (1/2,0) ellipse (0.75cm and 0.5cm);
\end{tikzpicture}},\widetilde{\mu}_{\begin{tikzpicture}[scale=0.6]
    \coordinate (A) at (0,0);
    \coordinate (B) at (1,0);
	\draw[thick] (A) to [bend left=60] (B);
    \draw[thick] (A) to [bend right=60] (B);
    \fill[black] (A) circle (2pt);
    \fill[black] (B) circle (2pt);
        \draw[thick, black] (0,0) ellipse (0.15cm and 0.15cm);
                \draw[thick, red] (1/2,0) ellipse (0.75cm and 0.5cm);
\end{tikzpicture}},\widetilde{\mu}_{\begin{tikzpicture}[scale=0.6]
    \coordinate (A) at (0,0);
    \coordinate (B) at (1,0);
	\draw[thick] (A) to [bend left=60] (B);
    \draw[thick] (A) to [bend right=60] (B);
    \fill[black] (A) circle (2pt);
    \fill[black] (B) circle (2pt);
        \draw[thick, black] (1,0) ellipse (0.15cm and 0.15cm);
                \draw[thick, red] (1/2,0) ellipse (0.75cm and 0.5cm);
\end{tikzpicture}},\widetilde{\mu}_{\input{Figures/graph_cyclic_2_tubing_12_m1.tex}},\widetilde{\mu}_{\input{Figures/graph_cyclic_2_tubing_1_m1.tex}},\widetilde{\mu}_{\input{Figures/graph_cyclic_2_tubing_2_m1.tex}},\widetilde{\mu}_{\input{Figures/graph_cyclic_2_tubing_12_m2.tex}},\widetilde{\mu}_{\input{Figures/graph_cyclic_2_tubing_1_m2.tex}},\widetilde{\mu}_{\input{Figures/graph_cyclic_2_tubing_2_m2.tex}},\widetilde{\mu}_{\input{Figures/graph_cyclic_1_1_tubing_1_2.tex}}\right\}\,,
\end{equation}
where
\begin{align}
\widetilde{\mu}_{\begin{tikzpicture}[scale=0.6]
    \coordinate (A) at (0,0);
    \coordinate (B) at (1,0);
	\draw[thick] (A) to [bend left=60] (B);
    \draw[thick] (A) to [bend right=60] (B);
    \fill[black] (A) circle (2pt);
    \fill[black] (B) circle (2pt);

                \draw[thick, red] (1/2,0) ellipse (0.75cm and 0.5cm);
\end{tikzpicture}}&=\Dlog \frac{x_1+X_1+x_2+X_2}{x_1}\wedge \Dlog \frac{x_1+X_1+x_2+X_2}{x_2}\,,\\
\widetilde{\mu}_{\begin{tikzpicture}[scale=0.6]
    \coordinate (A) at (0,0);
    \coordinate (B) at (1,0);
	\draw[thick] (A) to [bend left=60] (B);
    \draw[thick] (A) to [bend right=60] (B);
    \fill[black] (A) circle (2pt);
    \fill[black] (B) circle (2pt);
        \draw[thick, black] (0,0) ellipse (0.15cm and 0.15cm);
                \draw[thick, red] (1/2,0) ellipse (0.75cm and 0.5cm);
\end{tikzpicture}}&=\Dlog \frac{x_1+X_1+Y_{1,2}^{(1)}+Y_{1,2}^{(2)}}{x_1}\wedge \Dlog \frac{x_1+X_1+x_2+X_2}{x_2}\,,\\
\widetilde{\mu}_{\begin{tikzpicture}[scale=0.6]
    \coordinate (A) at (0,0);
    \coordinate (B) at (1,0);
	\draw[thick] (A) to [bend left=60] (B);
    \draw[thick] (A) to [bend right=60] (B);
    \fill[black] (A) circle (2pt);
    \fill[black] (B) circle (2pt);
        \draw[thick, black] (1,0) ellipse (0.15cm and 0.15cm);
                \draw[thick, red] (1/2,0) ellipse (0.75cm and 0.5cm);
\end{tikzpicture}}&=\Dlog \frac{x_1+X_1+x_2+X_2}{x_1}\wedge \Dlog \frac{x_2+X_2+Y_{1,2}^{(1)}+Y_{1,2}^{(2)}}{x_2}\,,\\
\widetilde{\mu}_{\input{Figures/graph_cyclic_2_tubing_12_m1.tex}}&=\Dlog \frac{x_1+X_1+x_2+X_2+2Y_{1,2}^{(1)}}{x_1}\wedge \Dlog \frac{x_1+X_1+x_2+X_2+2Y_{1,2}^{(1)}}{x_2}\,,\\
\widetilde{\mu}_{\input{Figures/graph_cyclic_2_tubing_1_m1.tex}}&=\Dlog \frac{x_1+X_1+Y_{1,2}^{(1)}+Y_{1,2}^{(2)}}{x_1}\wedge \Dlog \frac{x_1+X_1+x_2+X_2+2Y_{1,2}^{(1)}}{x_2}\,,\\
\widetilde{\mu}_{\input{Figures/graph_cyclic_2_tubing_2_m1.tex}}&=\Dlog \frac{x_1+X_1+x_2+X_2+2Y_{1,2}^{(1)}}{x_1}\wedge \Dlog \frac{x_2+X_2+Y_{1,2}^{(1)}+Y_{1,2}^{(2)}}{x_2}\,,\\
\widetilde{\mu}_{\input{Figures/graph_cyclic_2_tubing_12_m2.tex}}&=\Dlog \frac{x_1+X_1+x_2+X_2+2Y_{1,2}^{(2)}}{x_1}\wedge \Dlog \frac{x_1+X_1+x_2+X_2+2Y_{1,2}^{(2)}}{x_2}\,,\\
\widetilde{\mu}_{\input{Figures/graph_cyclic_2_tubing_1_m2.tex}}&=\Dlog \frac{x_1+X_1+Y_{1,2}^{(1)}+Y_{1,2}^{(2)}}{x_1}\wedge \Dlog \frac{x_1+X_1+x_2+X_2+2Y_{1,2}^{(2)}}{x_2}\,,\\
\widetilde{\mu}_{\input{Figures/graph_cyclic_2_tubing_2_m2.tex}}&=\Dlog \frac{x_1+X_1+x_2+X_2+2Y_{1,2}^{(2)}}{x_1}\wedge \Dlog \frac{x_2+X_2+Y_{1,2}^{(1)}+Y_{1,2}^{(2)}}{x_2}\,,\\
\widetilde{\mu}_{\input{Figures/graph_cyclic_1_1_tubing_1_2.tex}}&=\Dlog \frac{x_1+X_1+Y_{1,2}^{(1)}+Y_{1,2}^{(2)}}{x_1}\wedge \Dlog \frac{x_2+X_2+Y_{1,2}^{(1)}+Y_{1,2}^{(2)}}{x_2}\,.
\end{align}
The canonical form associated to the wavefunction can be written in terms of these differential forms as
\begin{align}\label{eq:cyclic2decomp}
\widetilde{\Omega}^{\text{flat}}_{\begin{tikzpicture}[scale=0.6]
    \coordinate (A) at (0,0);
    \coordinate (B) at (1,0);
	\draw[thick] (A) to [bend left=60] (B);
    \draw[thick] (A) to [bend right=60] (B);
    \fill[black] (A) circle (2pt);
    \fill[black] (B) circle (2pt);
\end{tikzpicture}}(X_v+x_v,Y_e)&=\left(
\widetilde{\mu}_{\begin{tikzpicture}[scale=0.6]
    \coordinate (A) at (0,0);
    \coordinate (B) at (1,0);
	\draw[thick] (A) to [bend left=60] (B);
    \draw[thick] (A) to [bend right=60] (B);
    \fill[black] (A) circle (2pt);
    \fill[black] (B) circle (2pt);
        \draw[thick, black] (0,0) ellipse (0.15cm and 0.15cm);
                \draw[thick, red] (1/2,0) ellipse (0.75cm and 0.5cm);
\end{tikzpicture}}+\widetilde{\mu}_{\begin{tikzpicture}[scale=0.6]
    \coordinate (A) at (0,0);
    \coordinate (B) at (1,0);
	\draw[thick] (A) to [bend left=60] (B);
    \draw[thick] (A) to [bend right=60] (B);
    \fill[black] (A) circle (2pt);
    \fill[black] (B) circle (2pt);
        \draw[thick, black] (1,0) ellipse (0.15cm and 0.15cm);
                \draw[thick, red] (1/2,0) ellipse (0.75cm and 0.5cm);
\end{tikzpicture}}-
\widetilde{\mu}_{\begin{tikzpicture}[scale=0.6]
    \coordinate (A) at (0,0);
    \coordinate (B) at (1,0);
	\draw[thick] (A) to [bend left=60] (B);
    \draw[thick] (A) to [bend right=60] (B);
    \fill[black] (A) circle (2pt);
    \fill[black] (B) circle (2pt);

                \draw[thick, red] (1/2,0) ellipse (0.75cm and 0.5cm);
\end{tikzpicture}}\right)-\left(
\widetilde{\mu}_{\input{Figures/graph_cyclic_2_tubing_1_m1.tex}}+\widetilde{\mu}_{\input{Figures/graph_cyclic_2_tubing_2_m1.tex}}-
\widetilde{\mu}_{\input{Figures/graph_cyclic_2_tubing_12_m1.tex}}\right)\nonumber\\
&-\left(
\widetilde{\mu}_{\input{Figures/graph_cyclic_2_tubing_1_m2.tex}}+\widetilde{\mu}_{\input{Figures/graph_cyclic_2_tubing_2_m2.tex}}-
\widetilde{\mu}_{\input{Figures/graph_cyclic_2_tubing_12_m2.tex}}\right)
+\widetilde{\mu}_{\input{Figures/graph_cyclic_1_1_tubing_1_2.tex}}\,,
\end{align}
where each bracket corresponds to a different product of amplitubes.
Geometrically, the relation \eqref{eq:cyclic2decomp} can be illustrated as in Fig.~\ref{fig:cyclic2}.

\begin{figure}[t]
\begin{center}
\input{Figures/graph_cyclic_2_regions.tex}
\end{center}
\caption{Geometric interpretation of the relation between $\widetilde{\Omega}$ and $\omega$ for the cyclic graph with two vertices.}
\label{fig:cyclic2}
\end{figure}

The action of the $\Dlog u_{\begin{tikzpicture}[scale=0.6]
    \coordinate (A) at (0,0);
    \coordinate (B) at (1,0);
	\draw[thick] (A) to [bend left=60] (B);
    \draw[thick] (A) to [bend right=60] (B);
    \fill[black] (A) circle (2pt);
    \fill[black] (B) circle (2pt);
\end{tikzpicture}}=\alpha_1\Dlog x_1+\alpha_2\Dlog x_2$
on each of the forms can be easily derived and is of the form
\begin{equation}
\Dlog u_{\begin{tikzpicture}[scale=0.6]
    \coordinate (A) at (0,0);
    \coordinate (B) at (1,0);
	\draw[thick] (A) to [bend left=60] (B);
    \draw[thick] (A) to [bend right=60] (B);
    \fill[black] (A) circle (2pt);
    \fill[black] (B) circle (2pt);
\end{tikzpicture}}\wedge v_{\begin{tikzpicture}[scale=0.6]
    \coordinate (A) at (0,0);
    \coordinate (B) at (1,0);
	\draw[thick] (A) to [bend left=60] (B);
    \draw[thick] (A) to [bend right=60] (B);
    \fill[black] (A) circle (2pt);
    \fill[black] (B) circle (2pt);
\end{tikzpicture}}= B_{\begin{tikzpicture}[scale=0.6]
    \coordinate (A) at (0,0);
    \coordinate (B) at (1,0);
	\draw[thick] (A) to [bend left=60] (B);
    \draw[thick] (A) to [bend right=60] (B);
    \fill[black] (A) circle (2pt);
    \fill[black] (B) circle (2pt);
\end{tikzpicture}}\wedge v_{\begin{tikzpicture}[scale=0.6]
    \coordinate (A) at (0,0);
    \coordinate (B) at (1,0);
	\draw[thick] (A) to [bend left=60] (B);
    \draw[thick] (A) to [bend right=60] (B);
    \fill[black] (A) circle (2pt);
    \fill[black] (B) circle (2pt);
\end{tikzpicture}}=\alpha_1 B_{\begin{tikzpicture}[scale=0.6]
    \coordinate (A) at (0,0);
    \coordinate (B) at (1,0);
	\draw[thick] (A) to [bend left=60] (B);
    \draw[thick] (A) to [bend right=60] (B);
    \fill[black] (A) circle (2pt);
    \fill[black] (B) circle (2pt);
\end{tikzpicture}}^{(1)}\wedge v_{\begin{tikzpicture}[scale=0.6]
    \coordinate (A) at (0,0);
    \coordinate (B) at (1,0);
	\draw[thick] (A) to [bend left=60] (B);
    \draw[thick] (A) to [bend right=60] (B);
    \fill[black] (A) circle (2pt);
    \fill[black] (B) circle (2pt);
\end{tikzpicture}}+\alpha_2 B_{\begin{tikzpicture}[scale=0.6]
    \coordinate (A) at (0,0);
    \coordinate (B) at (1,0);
	\draw[thick] (A) to [bend left=60] (B);
    \draw[thick] (A) to [bend right=60] (B);
    \fill[black] (A) circle (2pt);
    \fill[black] (B) circle (2pt);
\end{tikzpicture}}^{(2)}\wedge v_{\begin{tikzpicture}[scale=0.6]
    \coordinate (A) at (0,0);
    \coordinate (B) at (1,0);
	\draw[thick] (A) to [bend left=60] (B);
    \draw[thick] (A) to [bend right=60] (B);
    \fill[black] (A) circle (2pt);
    \fill[black] (B) circle (2pt);
\end{tikzpicture}}\,.
\end{equation}
The matrices $B^{(i)}_{\begin{tikzpicture}[scale=0.6]
    \coordinate (A) at (0,0);
    \coordinate (B) at (1,0);
	\draw[thick] (A) to [bend left=60] (B);
    \draw[thick] (A) to [bend right=60] (B);
    \fill[black] (A) circle (2pt);
    \fill[black] (B) circle (2pt);
\end{tikzpicture}}$ encoding this action are block diagonal with one $1\times 1$ block and three $3\times 3$ blocks. Since the three non-trivial blocks are very similar to each other, and similar to the path graph with two vertices we studied in the previous section, we just include one of them in detail:
\begin{equation}
B_{\input{Figures/graph_cyclic_2_tubing_12_m1.tex}}^{(1)}=\begin{pmatrix}
\Dlog R_{\input{Figures/graph_cyclic_2_region_12_m1.tex}}&0&0\\
0&\Dlog R_{\input{Figures/graph_cyclic_2_region_1_m1.tex}}&0\\
\Dlog\frac{R_{\input{Figures/graph_cyclic_2_region_12_m1.tex}}}{R_{\input{Figures/graph_cyclic_2_region_12_2_m1.tex}}}&0& \Dlog R_{\input{Figures/graph_cyclic_2_region_12_2_m1.tex}}
\end{pmatrix}
\end{equation}
and
\begin{equation}
B_{\input{Figures/graph_cyclic_2_tubing_12_m1.tex}}^{(2)}=\begin{pmatrix}
\Dlog R_{\input{Figures/graph_cyclic_2_region_12_m1.tex}}&0&0\\
\Dlog\frac{R_{\input{Figures/graph_cyclic_2_region_12_m1.tex}}}{R_{\input{Figures/graph_cyclic_2_region_12_1_m1.tex}}}&\Dlog R_{\input{Figures/graph_cyclic_2_region_12_1_m1.tex}}&0\\
0&0& \Dlog R_{\input{Figures/graph_cyclic_2_region_2_m1.tex}}
\end{pmatrix} \,,
\end{equation}
where we used the three canonical forms $v_{\input{Figures/graph_cyclic_2_tubing_12_m1.tex}}=\left\{\widetilde{\mu}_{\input{Figures/graph_cyclic_2_tubing_12_m1.tex}},\widetilde{\mu}_{\input{Figures/graph_cyclic_2_tubing_1_m1.tex}},\widetilde{\mu}_{\input{Figures/graph_cyclic_2_tubing_2_m1.tex}}\right\}$.

We can again diagonalise the coefficient matrix $B_{\begin{tikzpicture}[scale=0.6]
    \coordinate (A) at (0,0);
    \coordinate (B) at (1,0);
	\draw[thick] (A) to [bend left=60] (B);
    \draw[thick] (A) to [bend right=60] (B);
    \fill[black] (A) circle (2pt);
    \fill[black] (B) circle (2pt);
\end{tikzpicture}}$, and the diagonalisation process is independent in each block. For the $3\times 3$ block above, it leads to
\begin{align}
\overline{\mu}_{\input{Figures/graph_cyclic_2_tubing_12_m1.tex}}&=\widetilde{\mu}_{\input{Figures/graph_cyclic_2_tubing_12_m1.tex}}\,,\\
\overline{\mu}_{\input{Figures/graph_cyclic_2_tubing_1_m1.tex}}&=\left(\frac{\alpha_1(X_2+Y_{1,2}^{(1)}-Y_{1,2}^{(2)})}{\alpha_2(X_1+Y_{1,2}^{(1)}+Y_{1,2}^{(2)})}-1\right)\widetilde{\mu}_{\input{Figures/graph_cyclic_2_tubing_1_m1.tex}}+\widetilde{\mu}_{\input{Figures/graph_cyclic_2_tubing_12_m1.tex}}\,,\\
\overline{\mu}_{\input{Figures/graph_cyclic_2_tubing_2_m1.tex}}&=\left(\frac{\alpha_2(X_1+Y_{1,2}^{(1)}-Y_{1,2}^{(2)})}{\alpha_1(X_2+Y_{1,2}^{(1)}+Y_{1,2}^{(2)})}-1\right)\widetilde{\mu}_{\input{Figures/graph_cyclic_2_tubing_2_m1.tex}}+\widetilde{\mu}_{\input{Figures/graph_cyclic_2_tubing_12_m1.tex}}\,,
\end{align} 
with the diagonal action given as
\begin{align}
\Dlog u_{\begin{tikzpicture}[scale=0.6]
    \coordinate (A) at (0,0);
    \coordinate (B) at (1,0);
	\draw[thick] (A) to [bend left=60] (B);
    \draw[thick] (A) to [bend right=60] (B);
    \fill[black] (A) circle (2pt);
    \fill[black] (B) circle (2pt);
\end{tikzpicture}}\wedge \overline{\mu}_{\input{Figures/graph_cyclic_2_tubing_12_m1.tex}}&=(\alpha_1+\alpha_2)\Dlog(X_1+X_2+2Y_{1,2}^{(1)})\wedge\overline{\mu}_{\input{Figures/graph_cyclic_2_tubing_12_m1.tex}\,,}\\
\Dlog u_{\begin{tikzpicture}[scale=0.6]
    \coordinate (A) at (0,0);
    \coordinate (B) at (1,0);
	\draw[thick] (A) to [bend left=60] (B);
    \draw[thick] (A) to [bend right=60] (B);
    \fill[black] (A) circle (2pt);
    \fill[black] (B) circle (2pt);
\end{tikzpicture}}\wedge \overline{\mu}_{\input{Figures/graph_cyclic_2_tubing_1_m1.tex}}&=\left(\alpha_1\Dlog(X_1+Y_{1,2}^{(1)}+Y_{1,2}^{(2)})+\alpha_2 \Dlog(X_2+Y_{1,2}^{(1)}-Y_{1,2}^{(2)})\right)\wedge\overline{\mu}_{\input{Figures/graph_cyclic_2_tubing_1_m1.tex}}\,,\\
\Dlog u_{\begin{tikzpicture}[scale=0.6]
    \coordinate (A) at (0,0);
    \coordinate (B) at (1,0);
	\draw[thick] (A) to [bend left=60] (B);
    \draw[thick] (A) to [bend right=60] (B);
    \fill[black] (A) circle (2pt);
    \fill[black] (B) circle (2pt);
\end{tikzpicture}}\wedge \overline{\mu}_{\input{Figures/graph_cyclic_2_tubing_2_m1.tex}}&=\left(\alpha_1\Dlog(X_1+Y_{1,2}^{(1)}-Y_{1,2}^{(2)})+\alpha_2 \Dlog(X_2+Y_{1,2}^{(1)}+Y_{1,2}^{(2)})\right)\wedge\overline{\mu}_{\input{Figures/graph_cyclic_2_tubing_2_m1.tex}}\,,
\end{align}
where again each coefficient is just an appropriate region variable.

\section{Conclusions and Outlook}
In this paper we provided a new set of differential equations for wavefunction coefficients in power-law FRW cosmologies. While the motivation for these equations comes from physics, the solution beautifully combines algebra, geometry and combinatorics, and relies on the structure of the kinematic spaces of this physical problem. The equations that we derive rely on the notion of positive geometries, and the set of master functions that we propose is composed of twisted integrals of canonical forms of positive geometries. The fact that our integrands are logarithmic, up to the twist function, is fundamental in our derivation and allows us to derive a very general formula \eqref{eq:diffeq2}. Crucially, the functions that we introduce here have a very natural labelling from the point of view of the Feynman graphs in the perturbative decomposition of wavefunctions as the $c$-tubings of the graphs. As importantly, the coefficients of our differential equations are labelled by regions of these tubings, which allows us for a uniform combinatorial description of the system. The resulting formula \eqref{eq:finaleqs} is easy to check for any graph that one wants to consider. Finally, the diagonalisation process of the coefficient matrix of these differential equations can be performed explicitly and the resulting eigensystem, given by the eigenvalues and eigenvectors in \eqref{eq:eigenvalue} and \eqref{eq:eigenvector}, has again a very natural description from the point of view of the combinatorics of each graph.

While we believe that the equations we derived in this paper can be applied to any graph $G$, our conjecture is based on the exploration for graphs with a limited number of vertices. However, by studying these examples, we can already see general patterns emerging. A general proof for various formulae conjectured in the paper is yet to be provided, including formulae \eqref{eq:FRWastubings}, \eqref{eq:finaleqs}, \eqref{eq:eigenvector} and \eqref{eq:eigenvalue}. We believe that these proofs will be combinatorial in nature, and we leave them for future work.

The results in this paper provide a starting point for many other interesting questions.
The most urgent open problem is  how to solve the set of differential equations that we conjectured. Since our equations have a uniform description in terms of tubings on graph $G$, it would be interesting to explore if this combinatorial description also pertains to the solutions of these differential equations.
Moreover, all our considerations in this paper focused on the single Feynman diagrams in perturbation theory separately. Recently, a new method for combining various graphs contributing to the same wavefunction as a single positive geometry was realised in the definition of the cosmohedra \cite{Arkani-Hamed:2024jbp}, see also \cite{Glew:2025otn}. A natural question that arises is whether it is possible to write a (possibly simpler) set of differential equations starting from the canonical differential forms of the cosmohedra.

\section{Acknowledgements}

We would like to thank Ross Glew, Shounak De, Harry Goodhew, Andrzej Pokraka, Lorenzo Tancredi, Guilherme Pimentel and Stefan Weinzierl for discussions at various stages of this project.
This work was supported in part by the Deutsche Forschungsgemeinschaft (DFG, German Research Foundation) Projektnummer 508889767/FOR5582
``Modern Foundations of Scattering Amplitudes'', and by grant NSF PHY-2309135 to the Kavli Institute for Theoretical Physics (KITP).

\appendix


\bibliographystyle{nb}

\bibliography{cosmo}

\begin{thebibliography}{10}
\ifx\href\asklfhas\newcommand{\href}[2]{#2}\fi
\ifx\arxivref\asklfhas\newcommand{\arxivref}[2]{\href{http://arxiv.org/abs/#1}{#2}}\fi
\ifx\doiref\asklfhas\newcommand{\doiref}[2]{\href{http://dx.doi.org/#1}{#2}}\fi
\raggedright
\small
\parskip 0pt

\bibitem{Arkani-Hamed:2017tmz}
N.~Arkani-Hamed, Y.~Bai and T.~Lam,
\textit{``{Positive Geometries and Canonical Forms}''},
\textsf{\doiref{10.1007/JHEP11(2017)039}{JHEP~1711,~039~(2017)}},
\texttt{\arxivref{1703.04541}{arxiv:1703.04541}}.

\bibitem{Arkani-Hamed:2013jha}
N.~Arkani-Hamed and J.~Trnka,
\textit{``{The Amplituhedron}''},
\textsf{\doiref{10.1007/JHEP10(2014)030}{JHEP~1410,~030~(2014)}},
\texttt{\arxivref{1312.2007}{arxiv:1312.2007}}.

\bibitem{Damgaard:2019ztj}
D.~Damgaard, L.~Ferro, T.~Lukowski and M.~Parisi,
\textit{``{The Momentum Amplituhedron}''},
\textsf{\doiref{10.1007/JHEP08(2019)042}{JHEP~1908,~042~(2019)}},
\texttt{\arxivref{1905.04216}{arxiv:1905.04216}}.

\bibitem{Ferro:2022abq}
L.~Ferro and T.~Lukowski,
\textit{``{The Loop Momentum Amplituhedron}''},
\textsf{\doiref{10.1007/JHEP05(2023)183}{JHEP~2305,~183~(2023)}},
\texttt{\arxivref{2210.01127}{arxiv:2210.01127}}.

\bibitem{Arkani-Hamed:2017fdk}
N.~Arkani-Hamed, P.~Benincasa and A.~Postnikov,
\textit{``{Cosmological Polytopes and the Wavefunction of the Universe}''},
\texttt{\arxivref{1709.02813}{arxiv:1709.02813}}.

\bibitem{Arkani-Hamed:2018bjr}
N.~Arkani-Hamed and P.~Benincasa,
\textit{``{On the Emergence of Lorentz Invariance and Unitarity from the
  Scattering Facet of Cosmological Polytopes}''},
\texttt{\arxivref{1811.01125}{arxiv:1811.01125}}.

\bibitem{Benincasa:2024leu}
P.~Benincasa and G.~Dian,
\textit{``{The Geometry of Cosmological Correlators}''},
\textsf{\doiref{10.21468/SciPostPhys.18.3.105}{SciPost~Phys.~18,~105~(2025)}},
\texttt{\arxivref{2401.05207}{arxiv:2401.05207}}.

\bibitem{De:2023xue}
S.~De and A.~Pokraka,
\textit{``{Cosmology meets cohomology}''},
\textsf{\doiref{10.1007/JHEP03(2024)156}{JHEP~2403,~156~(2024)}},
\texttt{\arxivref{2308.03753}{arxiv:2308.03753}}.

\bibitem{De:2024zic}
S.~De and A.~Pokraka,
\textit{``{A physical basis for cosmological correlators from cuts}''},
\textsf{\doiref{10.1007/JHEP03(2025)040}{JHEP~2503,~040~(2025)}},
\texttt{\arxivref{2411.09695}{arxiv:2411.09695}}.

\bibitem{Gasparotto:2024bku}
F.~Gasparotto, P.~Mazloumi and X.~Xu,
\textit{``{Differential equations for tree-level cosmological correlators with
  massive states}''},
\textsf{\doiref{10.1007/JHEP09(2025)043}{JHEP~2509,~043~(2025)}},
\texttt{\arxivref{2411.05632}{arxiv:2411.05632}}.

\bibitem{Chen:2023iix}
J.~Chen and B.~Feng,
\textit{``{Towards systematic evaluation of de Sitter correlators via
  Generalized Integration-By-Parts relations}''},
\textsf{\doiref{10.1007/JHEP06(2024)199}{JHEP~2406,~199~(2024)}},
\texttt{\arxivref{2401.00129}{arxiv:2401.00129}}.

\bibitem{Chen:2024glu}
J.~Chen, B.~Feng and Y.-X.~Tao,
\textit{``{Multivariate hypergeometric solutions of cosmological (dS)
  correlators by d log-form differential equations}''},
\textsf{\doiref{10.1007/JHEP03(2025)075}{JHEP~2503,~075~(2025)}},
\texttt{\arxivref{2411.03088}{arxiv:2411.03088}}.

\bibitem{Arkani-Hamed:2023kig}
N.~Arkani-Hamed, D.~Baumann, A.~Hillman, A.~Joyce, H.~Lee and G.~L.~Pimentel,
\textit{``{Differential Equations for Cosmological Correlators}''},
\texttt{\arxivref{2312.05303}{arxiv:2312.05303}}.

\bibitem{Arkani-Hamed:2023bsv}
N.~Arkani-Hamed, D.~Baumann, A.~Hillman, A.~Joyce, H.~Lee and G.~L.~Pimentel,
\textit{``{Kinematic Flow and the Emergence of Time}''},
\texttt{\arxivref{2312.05300}{arxiv:2312.05300}}.

\bibitem{Baumann:2024mvm}
D.~Baumann, H.~Goodhew and H.~Lee,
\textit{``{Kinematic Flow for Cosmological Loop Integrands}''},
\texttt{\arxivref{2410.17994}{arxiv:2410.17994}}.

\bibitem{Baumann:2025qjx}
D.~Baumann, H.~Goodhew, A.~Joyce, H.~Lee, G.~L.~Pimentel and T.~Westerdijk,
\textit{``{Geometry of Kinematic Flow}''},
\texttt{\arxivref{2504.14890}{arxiv:2504.14890}}.

\bibitem{Fan:2024iek}
B.~Fan and Z.-Z.~Xianyu,
\textit{``{Cosmological amplitudes in power-law FRW universe}''},
\textsf{\doiref{10.1007/JHEP12(2024)042}{JHEP~2412,~042~(2024)}},
\texttt{\arxivref{2403.07050}{arxiv:2403.07050}}.

\bibitem{Grimm:2024mbw}
T.~W.~Grimm, A.~Hoefnagels and M.~van~Vliet,
\textit{``{Structure and complexity of cosmological correlators}''},
\textsf{\doiref{10.1103/PhysRevD.110.123531}{Phys.~Rev.~D~110,~123531~(2024)}},
\texttt{\arxivref{2404.03716}{arxiv:2404.03716}}.

\bibitem{Grimm:2024tbg}
T.~W.~Grimm and A.~Hoefnagels,
\textit{``{Reductions of GKZ systems and applications to cosmological
  correlators}''},
\textsf{\doiref{10.1007/JHEP04(2025)196}{JHEP~2504,~196~(2025)}},
\texttt{\arxivref{2409.13815}{arxiv:2409.13815}}.

\bibitem{Grimm:2025zhv}
T.~W.~Grimm, A.~Hoefnagels and M.~van~Vliet,
\textit{``{A Reduction Algorithm for Cosmological Correlators: Cuts,
  Contractions, and Complexity}''},
\texttt{\arxivref{2503.05866}{arxiv:2503.05866}}.

\bibitem{He:2024olr}
S.~He, X.~Jiang, J.~Liu, Q.~Yang and Y.-Q.~Zhang,
\textit{``{Differential equations and recursive solutions for cosmological
  amplitudes}''},
\textsf{\doiref{10.1007/JHEP01(2025)001}{JHEP~2501,~001~(2025)}},
\texttt{\arxivref{2407.17715}{arxiv:2407.17715}}.

\bibitem{Benincasa:2024ptf}
P.~Benincasa, G.~Brunello, M.~K.~Mandal, P.~Mastrolia and F.~Vaz\~ao,
\textit{``{One-loop corrections to the Bunch-Davies wave function of the
  universe}''},
\textsf{\doiref{10.1103/PhysRevD.111.085016}{Phys.~Rev.~D~111,~085016~(2025)}},
\texttt{\arxivref{2408.16386}{arxiv:2408.16386}}.

\bibitem{Hang:2024xas}
Y.~Hang and C.~Shen,
\textit{``{A Note on Kinematic Flow and Differential Equations for Two-Site
  One-Loop Graph in FRW Spacetime}''},
\texttt{\arxivref{2410.17192}{arxiv:2410.17192}}.

\bibitem{Fevola:2024nzj}
C.~Fevola, G.~L.~Pimentel, A.-L.~Sattelberger and T.~Westerdijk,
\textit{``{Algebraic Approaches to Cosmological Integrals}''},
\texttt{\arxivref{2410.14757}{arxiv:2410.14757}}.

\bibitem{Glew:2025ugf}
R.~Glew,
\textit{``{Wavefunction coefficients from Amplitubes}''},
\texttt{\arxivref{2503.13596}{arxiv:2503.13596}}.

\bibitem{Glew:2025otn}
R.~Glew and T.~Lukowski,
\textit{``{Amplitubes: Graph Cosmohedra}''},
\texttt{\arxivref{2502.17564}{arxiv:2502.17564}}.

\bibitem{CARR20062155}
M.~Carr and S.~L.~Devadoss,
\textit{``Coxeter complexes and graph-associahedra''},
\textsf{\doiref{https://doi.org/10.1016/j.topol.2005.08.010}{Topology~and~its~Applications~153,~2155~(2006)}}.

\bibitem{Henn:2013pwa}
J.~M.~Henn,
\textit{``{Multiloop integrals in dimensional regularization made simple}''},
\textsf{\doiref{10.1103/PhysRevLett.110.251601}{Phys.~Rev.~Lett.~110,~251601~(2013)}},
\texttt{\arxivref{1304.1806}{arxiv:1304.1806}}.

\bibitem{Arkani-Hamed:2024jbp}
N.~Arkani-Hamed, C.~Figueiredo and F.~Vaz\~ao,
\textit{``{Cosmohedra}''},
\texttt{\arxivref{2412.19881}{arxiv:2412.19881}}.

\end{thebibliography}

\end{document}